\documentclass[aps,prx,nofootinbib,superscriptaddress,twocolumn]{revtex4-2}

\usepackage{bookmark}
\usepackage{xcolor}
\hypersetup{colorlinks}

\usepackage{amsmath}
\usepackage{graphicx}
\usepackage{subcaption}
\usepackage[noabbrev]{cleveref}
\usepackage{tabularray}
\UseTblrLibrary{booktabs}   
\usepackage{makecell}  
\usepackage{amsthm,thmtools}
\theoremstyle{definition}
\newtheorem{theorem}{Theorem}

\usepackage{stmaryrd}
\SetSymbolFont{stmry}{bold}{U}{stmry}{m}{n}

\usepackage[bottom]{footmisc}

\usepackage{macros/tikzfig}

\tikzstyle{Z dot}=[inner sep=0mm, minimum size=2mm, shape=circle, draw=black, fill=zxGreen, tikzit fill={rgb,255: red,216; green,248; blue,216}, outer sep=-0.5mm]
\tikzstyle{Z phase dot}=[draw=black, fill=zxGreen, shape=rectangle, minimum size=4.5mm, rounded corners=1.8mm, inner sep=0.5mm, outer sep=-0.5mm, scale=0.8, tikzit shape=circle, tikzit fill={rgb,255: red,216; green,248; blue,216}, font={\footnotesize\boldmath}]
\tikzstyle{X dot}=[shape=circle, draw=black, fill=zxRed, tikzit fill={rgb,255: red,221; green,165; blue,165}, inner sep=0 mm, minimum size=2 mm, outer sep=-0.5mm]
\tikzstyle{X phase dot}=[Z phase dot, draw=black, fill=zxRed, tikzit fill={rgb,255: red,221; green,165; blue,165}]
\tikzstyle{hadamard}=[fill=zxHad, draw=black, shape=rectangle, inner sep=0.6mm, minimum height=1.5mm, minimum width=1.5mm, tikzit fill=yellow, font={\scriptsize}]
\tikzstyle{box}=[draw=black, shape=rectangle, fill=white, minimum size=1em, inner sep=0.2em, scale=0.85, font={\scriptsize}, outer sep=-0.5mm]
\tikzstyle{black dot}=[fill=black, draw=black, shape=circle, inner sep=1pt, outer sep=-3pt]
\tikzstyle{white dot}=[fill=white, draw=black, shape=circle, inner sep=1.5pt, outer sep=-3pt]
\tikzstyle{sLabel}=[font={\scriptsize}, tikzit draw=black, auto]
\tikzstyle{floatingLabel}=[rectangle, fill=white, inner sep=0pt, font={\scriptsize}, label distance=1mm, fill opacity=.5, text opacity=1]
\tikzstyle{not}=[draw=black, circle, addcross, minimum size=2mm, outer sep=-3pt, inner sep=0mm]
\tikzstyle{meter}=[draw=black, fill=white, shape=rectangle, addmeter]
\tikzstyle{Z dot thick}=[inner sep=0mm, minimum size=2mm, shape=circle, draw=black, fill=zxGreen, tikzit fill={rgb,255: red,216; green,248; blue,216}, outer sep=-0.5mm, line width=1pt]
\tikzstyle{X dot thick}=[inner sep=0mm, minimum size=2mm, shape=circle, draw=black, fill=zxRed, tikzit fill={rgb,255: red,221; green,165; blue,165}, outer sep=-0.5mm, line width=1pt]
\tikzstyle{fault-location}=[fill=white, draw=black, shape=circle, minimum size=2mm, inner sep=0mm, outer sep=-0.5 mm, regular polygon, regular polygon sides=8]
\tikzstyle{fault-location-faulty}=[fill=white, draw={rgb,255: red,191; green,0; blue,64}, shape=circle, minimum size=2mm, inner sep=0mm, outer sep=-0.5 mm, regular polygon, regular polygon sides=8, minimum size=3mm]
\tikzstyle{new style 0}=[fill=white, draw=black, shape=circle]

\tikzstyle{dashed-line}=[-, style=dashed, draw={rgb,255: red,128; green,128; blue,128}]
\tikzstyle{hadamard edge}=[-, style=dashed, draw=blue]
\tikzstyle{X Web}=[-, preaction={line width=1mm, draw=zxDarkRed}, tikzit draw=red]
\tikzstyle{Z Web}=[-, preaction={line width=1.5mm, draw=zxDarkGreen}, tikzit draw=green]
\tikzstyle{error}=[-, draw=red, thick]
\tikzstyle{XZ Web}=[-, preaction={line width=1.8mm, draw=zxDarkGreen}, preaction={line width=1mm, draw=zxDarkRed}, tikzit draw=blue]
\tikzstyle{braceedge}=[-, decorate, decoration={brace, amplitude=2mm, raise=-1mm}]
\tikzstyle{arrow}=[->]
\tikzstyle{fault-free}=[-, draw={rgb,255: red,177; green,98; blue,255}, line width=1pt]
\tikzstyle{new edge style 0}=[-, fill={rgb,255: red,211; green,211; blue,211}, draw=none]
\tikzstyle{XerrorPropagation}=[-, draw=red, line width=1pt]
\tikzstyle{XerrorPropagationArrow}=[->, draw=red, line width=1pt]
\tikzstyle{ZerrorPropagation}=[-, draw=green, line width=1pt]
\tikzstyle{ZerrorPropagationArrow}=[->, draw=green, line width=1pt]
\tikzstyle{fault-accounting-component}=[-, style=dashed, draw={rgb,255: red,0; green,128; blue,128}]
\tikzstyle{new edge style 2}=[-, style=dashed, draw={rgb,255: red,0; green,128; blue,128}]

\input{macros/floquet.tikzdefs}

\newcommand{\code}[1]{\left\llbracket#1\right\rrbracket}
\newcommand{\pPhis}{p_{\mathrm{phys}}}
\newcommand{\pL}{p_{\mathrm{L}}}
\newcommand{\pMem}{p_{\mathrm{mem}}}
\newcommand{\pTwo}{p_{\mathrm{2}}}

\newcommand{\pSPAM}{p_{\mathrm{SPAM}}}

\begin{document}

\title{Ultra Low Overhead Syndrome Extraction for the Steane code}

\author{Boldizsár Poór}%
\thanks{Corresponding author}
\email{boldizsar.poor@gmail.com}
\affiliation{University of Oxford, United Kingdom}
\affiliation{Quantinuum, United Kingdom}
\author{Benjamin Rodatz}
\affiliation{University of Oxford, United Kingdom}
\affiliation{Quantinuum, United Kingdom}
\author{Aleks Kissinger}
\affiliation{University of Oxford, United Kingdom}


\begin{abstract}
  We establish a new performance benchmark for the fault-tolerant syndrome extraction of the $\code{7, 1, 3}$ Steane code with a dynamic protocol.
  Our method is built on two highly optimized circuits derived using fault-equivalent ZX-rewrites:
  a primary fault-tolerant circuit with 14 CNOTs and an efficient non-fault-tolerant recovery circuit with 11 CNOTs.
  The protocol uses an adaptive response to internal faults, discarding flagged measurements and falling back to the recovery circuit to correct potentially detrimental errors.
  Monte Carlo simulations confirm the efficiency of our protocol, reducing the logical error rate per cycle by an average of $\sim\! 14.3\%$ relative to the optimized Steane method~\cite{rodatzFaultTolerance2025} and $\sim\! 17.3\%$ compared to Reichardt's three-qubit method~\cite{reichardtFaulttolerantQuantum2020}, the leading prior techniques.
\end{abstract}

\keywords{QEC, FTQC, Steane Code, Fault Tolerance, Syndrome Measurement}

\maketitle

\section{Introduction}

The realization of a scalable quantum computer heavily relies on the implementation of fault-tolerant quantum error correction to protect quantum information from environmental noise.
In particular, one of the most ubiquitously used gadgets when aiming to realize fault-tolerant quantum computing (FTQC) is the syndrome measurement circuit.
Not only is it used to measure syndromes and correct errors regularly~\cite{shorFaulttolerantQuantum1996}, but it is also essential for many magic state preparation schemes~\cite{bravyiQuantumCodesLattice1998,chamberlandVeryLow2020,gidneyMagicState2024}.
It is therefore crucial that the implementation of this gadget is fault-tolerant and introduces minimal additional noise.

Within the vast space of quantum error correction (QEC) codes, the $\code{7, 1, 3}$ Steane code is one of the most fundamental and well-studied examples.
Discovered by Andrew Steane in 1996~\cite{steaneErrorCorrecting1996}, it is a self-dual CSS code constructed from the classical $[7, 4, 3]$ Hamming code, and is the smallest CSS code capable of correcting a single-qubit error~\cite{shawEncodingOne2008}.
As a CSS code, its X and Z syndromes can be extracted independently, making low overhead Steane-style syndrome extraction possible~\cite{steaneActiveStabilization1997}.
The recursive concatenation of the Steane code yields one of the first code families proven to admit a concatenated threshold~\cite{knillResilientQuantum1998}.
Due to these important theoretical properties and its small size, it remains one of the most experimentally realized QEC codes to date~\cite{ryan-andersonRealizationRealTime2021,paetznickDemonstrationLogical2024,bluvsteinLogicalQuantum2024,postlerDemonstrationFaultTolerant2024,daguerreExperimentalDemonstration2025,bluvsteinFaulttolerantNeutralatom2026,perlinFaulttolerantExecution2026}.

Beyond the generally applicable methods proposed by \citet{shorFaulttolerantQuantum1996} and \citet{steaneActiveStabilization1997}, several fault-tolerant (FT) syndrome extraction (SE) schemes have been developed for the Steane code.
For low overhead, the dominant strategy has been the dynamic `flag-and-fallback' approach.
Examples of this approach include the \emph{two-qubit} method proposed by~\citet{chaoQuantumError2018}, later improved by \citet{bhatnagarLowDepthFlagStyle2023}, and the \emph{three-qubit} method by \citet{reichardtFaulttolerantQuantum2020}.
This approach works by continuously measuring syndromes with a flagged circuit that can detect internal faults~\cite{chamberlandFlagFaulttolerant2018,chaoFlagFaultTolerant2020}.
If a flag is raised, the protocol switches to a simpler, non-fault-tolerant (non-FT) `recovery' SE circuit that diagnoses the most likely fault.
While the protocol includes a non-FT component, the full protocol is fault-tolerant according to the extended rectangle formalism for distance-three codes~\cite{aliferisQuantumAccuracy2006,chaoQuantumError2018}.

Recently, \citet{rodatzFaultTolerance2025} introduced a novel static SE circuit which, while not universally optimal, achieves state-of-the-art performance in many important contexts.
The circuit was derived using the ZX-calculus~\cite{coeckeInteractingQuantumObservables2008} --- a graphical language for representing and reasoning about quantum computation.
This was achieved by constructing a Steane-style SE circuit with the fault-tolerant state preparation circuit of \citet{gotoMinimizingResource2016}, and then reducing gate and qubit count using fault-equivalent rewrites~\cite{rodatzFaultTolerance2025,rodatzFloquetifyingStabiliser2024}.
While this static approach results in a highly efficient circuit, it can leave the logical state corrupted between syndrome measurements.

This work builds on the rewrite-based approach of~\citet{rodatzFaultTolerance2025}, introducing the following key improvements:
\begin{description}
    \item[Highly optimized extraction circuits] The derivation of a primary fault-tolerant circuit and a non-FT recovery circuit with reduced gate and ancilla counts.
    \item[Tight overhead bounds] The establishment of strict CNOT lower bounds for specific syndrome extraction circuits for the Steane code, proving that our circuits are optimal in their respective settings (see \Cref{prop:non-ft-cnot-optimality,prop:ft-cnot-optimality} for formal proofs).
    \item[Self-sufficient primary extraction] An adaptive protocol where the primary circuit is entirely self-sufficient in the no-flag case, requiring fallback only upon a raised flag, whereas previous protocols require the recovery circuit upon measuring any non-trivial syndrome.
\end{description}
Monte Carlo simulations under a realistic near-term hardware noise model demonstrate the high efficiency of our protocol.
It reduces the logical error rate per cycle by an average (over X and Z bases and the simulated number of cycles) of $\sim 14.3\%$ against the optimized Steane~\cite{rodatzFaultTolerance2025} and by $\sim 17.3\%$ against the three-qubit method~\cite{reichardtFaulttolerantQuantum2020}.

\section{Circuits and Dynamic Protocol}

In this section, we introduce our dynamic protocol that we call the \emph{Dynamic Optimized Steane} method, and the circuits that it uses.
The first circuit it uses is a fault-tolerant SE circuit optimized for low overhead, and the second is an even leaner non-fault-tolerant `recovery' circuit.
\Cref{fig:circuits} shows the final form of our circuits for measuring the Z stabilizers of the Steane code.

\begin{figure}
  \scriptsize
  \[
    \scalebox{.95}{\begin{tikzpicture}
	\begin{pgfonlayer}{nodelayer}
		\node [style=white dot] (0) at (0, 2) {};
		\node [style=white dot] (1) at (-1.25, 0) {};
		\node [style=white dot] (2) at (-2.5, -2) {};
		\node [style=white dot] (3) at (2.5, -2) {};
		\node [style=white dot] (4) at (1.25, 0) {};
		\node [style=white dot] (5) at (0, -0.75) {};
		\node [style=white dot] (6) at (0, -2) {};
		\node [style=sLabel] (7) at (-0.25, 2.375) {1};
		\node [style=sLabel] (8) at (-1.625, 0.25) {2};
		\node [style=sLabel] (9) at (-2.5, -2.375) {5};
		\node [style=sLabel] (10) at (1.625, 0.25) {4};
		\node [style=sLabel] (11) at (0, -2.375) {6};
		\node [style=sLabel] (12) at (2.5, -2.375) {7};
		\node [style=sLabel] (13) at (0, -0.25) {3};
	\end{pgfonlayer}
	\begin{pgfonlayer}{edgelayer}
		\draw [fill=green] (4.center)
			 to (5.center)
			 to (1.center)
			 to (0.center)
			 to [in=120, out=-60] cycle;
		\draw [fill=blue] (2.center)
			 to (6.center)
			 to (5.center)
			 to (1.center)
			 to cycle;
		\draw [fill=red] (5.center)
			 to (6.center)
			 to (3.center)
			 to (4.center)
			 to cycle;
	\end{pgfonlayer}
\end{tikzpicture}}
    \quad
    {
      \begin{array}{c|ccccccc}
        {\color{green} S_1} & X_1 & X_2 & X_3 & X_4 & I & I & I \\[2pt]
        {\color{blue} S_2} & I & X_2 & X_3 & I & X_5 & X_6 & I \\[2pt]
        {\color{red} S_3} & I & I & X_3 & X_4 & I & X_6 & X_7 \\[2pt] \hline \\[-5pt]
        {\color{green} S_4} & Z_1 & Z_2 & Z_3 & Z_4 & I & I & I \\[2pt]
        {\color{blue} S_5} & I & Z_2 & Z_3 & I & Z_5 & Z_6 & I \\[2pt]
        {\color{red} S_6} & I & I & Z_3 & Z_4 & I & Z_6 & Z_7 \\[2pt]
      \end{array}
    }
  \]
  \caption{Stabilizers of the Steane code~\cite{steaneErrorCorrecting1996}, with qubit-layout convention adopted from Ref.~\cite{ryan-andersonRealizationRealTime2021}.}
  \label{fig:qubit-layout}
\end{figure}

To derive these circuits, we start from the same circuit as in previous works~\cite{rodatzFaultTolerance2025}:
the Steane-style syndrome extraction circuit with Goto's probabilistic state preparation~\cite{gotoMinimizingResource2016} (adapted to the qubit layout shown in \Cref{fig:qubit-layout}).
Then, to reduce gate and qubit count, we apply fault-equivalent rewrites --- ZX manipulations that provably preserve a circuit's behaviour under noise, and thus, its tolerance to faults~\cite{rodatzFloquetifyingStabiliser2024,rodatzFaultTolerance2025}.
The non-FT version has its flag qubit removed to minimize overhead.
The full derivations are detailed in Appendix~\ref{sec:derivations}.

\begin{figure}
  \captionsetup[subfigure]{justification=raggedright,singlelinecheck=false,margin=0pt,aboveskip=-20pt}
  \begin{subfigure}[t]{\linewidth}
    \[
      \begin{tikzpicture}
	\begin{pgfonlayer}{nodelayer}
		\node [style=not] (0) at (3.25, -2) {};
		\node [style=not] (1) at (4, -3) {};
		\node [style=not] (2) at (-5.5, -3) {};
		\node [style=not] (3) at (4.75, -4) {};
		\node [style=not] (4) at (-4.75, -4) {};
		\node [style=not] (5) at (-3.75, -2) {};
		\node [style=none] (6) at (-6.25, -3) {};
		\node [style=none] (7) at (-6.25, -4) {};
		\node [style=none] (8) at (-6.25, -2) {};
		\node [style=black dot] (9) at (0.25, 3.5) {};
		\node [style=black dot] (10) at (3.25, 2.75) {};
		\node [style=black dot] (11) at (4, 2) {};
		\node [style=black dot] (12) at (-5.5, 1.25) {};
		\node [style=black dot] (13) at (4.75, 0.5) {};
		\node [style=black dot] (14) at (-4.75, -0.25) {};
		\node [style=black dot] (15) at (-3.75, -1) {};
		\node [style=none] (16) at (6, 3.5) {};
		\node [style=none] (17) at (6, 2.75) {};
		\node [style=none] (18) at (6, 2) {};
		\node [style=none] (19) at (6, 1.25) {};
		\node [style=none] (20) at (6, 0.5) {};
		\node [style=none] (21) at (6, -0.25) {};
		\node [style=none] (22) at (6, -1) {};
		\node [style=none] (23) at (-6.25, 3.5) {};
		\node [style=none] (24) at (-6.25, 2.75) {};
		\node [style=none] (25) at (-6.25, 2) {};
		\node [style=none] (26) at (-6.25, 1.25) {};
		\node [style=none] (27) at (-6.25, 0.5) {};
		\node [style=none] (28) at (-6.25, -0.25) {};
		\node [style=none] (29) at (-6.25, -1) {};
		\node [style=black dot] (30) at (1.25, -2) {};
		\node [style=not] (31) at (1.25, -4) {};
		\node [style=black dot] (32) at (-1.75, -4) {};
		\node [style=not] (33) at (-1.75, -2) {};
		\node [style=black dot] (34) at (-0.75, -2) {};
		\node [style=not] (35) at (0.25, -2) {};
		\node [style=not] (36) at (0.25, -4) {};
		\node [style=black dot] (37) at (0.25, -3) {};
		\node [style=not] (38) at (-0.75, -3) {};
		\node [style=sLabel] (39) at (-6.75, -2) {$\ket{0}$};
		\node [style=sLabel] (40) at (-6.75, -3) {$\ket{0}$};
		\node [style=sLabel] (41) at (-6.75, -4) {$\ket{0}$};
		\node [style=meter] (42) at (5.75, -3) {};
		\node [style=meter] (43) at (5.75, -2) {};
		\node [style=meter] (44) at (5.75, -4) {};
		\node [style=none] (45) at (-6.25, -5) {};
		\node [style=sLabel] (46) at (-6.75, -5) {$\ket{0}$};
		\node [style=box] (47) at (-5.5, -5) {$H$};
		\node [style=meter] (48) at (5.75, -5) {};
		\node [style=box] (49) at (4.75, -5) {$H$};
		\node [style=not] (50) at (-2.75, -3) {};
		\node [style=not] (51) at (-3.75, -4) {};
		\node [style=not] (52) at (2.25, -2) {};
		\node [style=black dot] (53) at (2.25, -5) {};
		\node [style=black dot] (54) at (-2.75, -5) {};
		\node [style=black dot] (55) at (-3.75, -5) {};
	\end{pgfonlayer}
	\begin{pgfonlayer}{edgelayer}
		\draw (8.center) to (5);
		\draw (6.center) to (2);
		\draw (7.center) to (4);
		\draw (15) to (5);
		\draw (4) to (14);
		\draw (3) to (13);
		\draw (12) to (2);
		\draw (1) to (11);
		\draw (10) to (0);
		\draw (29.center) to (15);
		\draw (15) to (22.center);
		\draw (21.center) to (14);
		\draw (14) to (28.center);
		\draw (27.center) to (13);
		\draw (13) to (20.center);
		\draw (19.center) to (12);
		\draw (12) to (26.center);
		\draw (25.center) to (11);
		\draw (11) to (18.center);
		\draw (17.center) to (10);
		\draw (10) to (24.center);
		\draw (23.center) to (9);
		\draw (16.center) to (9);
		\draw (32) to (33);
		\draw (34) to (35);
		\draw (30) to (31);
		\draw (37) to (38);
		\draw (38) to (34);
		\draw (31) to (36);
		\draw (5) to (33);
		\draw (3) to (31);
		\draw (1) to (37);
		\draw (35) to (9);
		\draw (35) to (30);
		\draw (33) to (34);
		\draw (36) to (37);
		\draw (32) to (36);
		\draw (0) to (43);
		\draw (1) to (42);
		\draw (44) to (3);
		\draw (45.center) to (47);
		\draw (48) to (49);
		\draw (53) to (52);
		\draw (0) to (52);
		\draw (52) to (30);
		\draw (49) to (53);
		\draw (47) to (55);
		\draw (55) to (54);
		\draw (54) to (53);
		\draw (54) to (50);
		\draw (51) to (55);
		\draw (4) to (51);
		\draw (32) to (51);
		\draw (2) to (50);
		\draw (50) to (38);
	\end{pgfonlayer}
\end{tikzpicture}
    \]
    \caption{}\label{fig:FT-circuit}
  \end{subfigure}
  \begin{subfigure}[t]{\linewidth}
    \[
      \begin{tikzpicture}
	\begin{pgfonlayer}{nodelayer}
		\node [style=not] (0) at (2.25, -2) {};
		\node [style=not] (1) at (3, -3) {};
		\node [style=not] (2) at (-4.25, -3) {};
		\node [style=not] (3) at (3.75, -4) {};
		\node [style=not] (4) at (-3.5, -4) {};
		\node [style=not] (5) at (-2.75, -2) {};
		\node [style=none] (6) at (-5, -3) {};
		\node [style=none] (7) at (-5, -4) {};
		\node [style=none] (8) at (-5, -2) {};
		\node [style=black dot] (9) at (0.25, 3.5) {};
		\node [style=black dot] (10) at (2.25, 2.75) {};
		\node [style=black dot] (11) at (3, 2) {};
		\node [style=black dot] (12) at (-4.25, 1.25) {};
		\node [style=black dot] (13) at (3.75, 0.5) {};
		\node [style=black dot] (14) at (-3.5, -0.25) {};
		\node [style=black dot] (15) at (-2.75, -1) {};
		\node [style=none] (16) at (5, 3.5) {};
		\node [style=none] (17) at (5, 2.75) {};
		\node [style=none] (18) at (5, 2) {};
		\node [style=none] (19) at (5, 1.25) {};
		\node [style=none] (20) at (5, 0.5) {};
		\node [style=none] (21) at (5, -0.25) {};
		\node [style=none] (22) at (5, -1) {};
		\node [style=none] (23) at (-5, 3.5) {};
		\node [style=none] (24) at (-5, 2.75) {};
		\node [style=none] (25) at (-5, 2) {};
		\node [style=none] (26) at (-5, 1.25) {};
		\node [style=none] (27) at (-5, 0.5) {};
		\node [style=none] (28) at (-5, -0.25) {};
		\node [style=none] (29) at (-5, -1) {};
		\node [style=black dot] (30) at (1.25, -2) {};
		\node [style=not] (31) at (1.25, -4) {};
		\node [style=black dot] (32) at (-1.75, -4) {};
		\node [style=not] (33) at (-1.75, -2) {};
		\node [style=black dot] (34) at (-0.75, -2) {};
		\node [style=not] (35) at (0.25, -2) {};
		\node [style=not] (36) at (0.25, -4) {};
		\node [style=black dot] (37) at (0.25, -3) {};
		\node [style=not] (38) at (-0.75, -3) {};
		\node [style=sLabel] (39) at (-5.5, -2) {$\ket{0}$};
		\node [style=sLabel] (40) at (-5.5, -3) {$\ket{0}$};
		\node [style=sLabel] (41) at (-5.5, -4) {$\ket{0}$};
		\node [style=meter] (45) at (4.75, -3) {};
		\node [style=meter] (46) at (4.75, -2) {};
		\node [style=meter] (47) at (4.75, -4) {};
	\end{pgfonlayer}
	\begin{pgfonlayer}{edgelayer}
		\draw (8.center) to (5);
		\draw (6.center) to (2);
		\draw (7.center) to (4);
		\draw (15) to (5);
		\draw (4) to (14);
		\draw (3) to (13);
		\draw (12) to (2);
		\draw (1) to (11);
		\draw (10) to (0);
		\draw (29.center) to (15);
		\draw (15) to (22.center);
		\draw (21.center) to (14);
		\draw (14) to (28.center);
		\draw (27.center) to (13);
		\draw (13) to (20.center);
		\draw (19.center) to (12);
		\draw (12) to (26.center);
		\draw (25.center) to (11);
		\draw (11) to (18.center);
		\draw (17.center) to (10);
		\draw (10) to (24.center);
		\draw (23.center) to (9);
		\draw (16.center) to (9);
		\draw (32) to (33);
		\draw (34) to (35);
		\draw (30) to (31);
		\draw (37) to (38);
		\draw (38) to (34);
		\draw (31) to (36);
		\draw (5) to (33);
		\draw (3) to (31);
		\draw (1) to (37);
		\draw (30) to (0);
		\draw (38) to (2);
		\draw (4) to (32);
		\draw (35) to (9);
		\draw (35) to (30);
		\draw (33) to (34);
		\draw (36) to (37);
		\draw (32) to (36);
		\draw (0) to (46);
		\draw (1) to (45);
		\draw (47) to (3);
	\end{pgfonlayer}
\end{tikzpicture}
    \]
    \caption{}\label{fig:non-FT-circuit}
  \end{subfigure}
  \caption{
    Optimized Z-syndrome extraction circuits of our dynamic protocol:
    (a) fault-tolerant \emph{Primary Z-SE Circuit} using 14 CNOTs and 4 ancillae, and
    (b) non-fault-tolerant \emph{Recovery Z-SE Circuit} using 11 CNOTs and 3 ancillae.
    The Steane code's self-duality yields structurally identical X-syndrome measurements.
  }
  \label{fig:circuits}
\end{figure}

\subsection{CNOT optimality}\label{subsec:cnot-optimality}

To formally establish the low resource overhead of these constructions, we present the following bounds.
\begin{restatable}[Non-FT CNOT Optimality]{proposition}{NonFtCnotOptimality}
  \label{prop:non-ft-cnot-optimality}
  For a non-destructive (and non-fault-tolerant) measurement of the three Z-syndromes of the Steane code, using three Z basis ancilla qubits and CNOT operations between data$\to$ancilla and ancilla$\to$ancilla, the minimal number of CNOT gates required is $11$.
\end{restatable}
\noindent\textit{Proof sketch.} This problem can be mapped to finding the shortest path in a finite state space of binary matrices representing the ancilla states.
Starting from the zero matrix, allowed CNOT operations correspond to directed edges.
An exhaustive breadth-first search confirms that no path of length $\le 10$ reaches the target Steane parity-check matrix, establishing the strict lower bound of 11; see Appendix~\ref{sec:cnot-optimal} for the full proof.

\begin{restatable}[FT Flagging Optimality]{proposition}{FtCnotOptimality}
  \label{prop:ft-cnot-optimality}
  Let $C$ be any non-destructive, CNOT-only circuit that measures the three Steane $Z$-syndromes using three ancilla qubits and exactly $11$~CNOT gates.
  Then $C$ must be augmented with at least three additional CNOT gates in order to flag (on a separate ancilla) every single-qubit ancilla fault that can propagate to the data as a weight-two $Z$ error.
\end{restatable}
\noindent\textit{Proof sketch.} By exhaustively simulating Pauli error propagation for all valid 11-CNOT circuits, we identify all dangerous internal faults.
A subsequent exhaustive search over all possible CNOT augmentations confirms that appending one or two CNOTs is insufficient to flag all such faults;
at least three additional CNOTs are strictly necessary; see Appendix~\ref{sec:cnot-optimal} for the full proof.

\subsection{Protocol Operation}

The operational flow of our dynamic protocol is summarized in the flowchart in \Cref{fig:full-protocol}.
A cycle begins with the execution of the primary FT circuit (\Cref{fig:FT-circuit}).
If the flag qubit measurement outcome is $0$, the results are decoded using the standard procedure detailed subsequently in \Cref{subsubsec:no-flag-raised}.
If the flag outcome is $1$, the measurement outputs from the primary SE gadget are discarded.
The protocol then executes the non-FT `recovery' circuit (\Cref{fig:non-FT-circuit}) in the dual basis, and the measurement outcomes from this recovery run are decoded using the modified logic described in \Cref{subsubsec:flag-raised}.

\begin{figure}[bt]
  \[
    \scalebox{.75}{\begin{tikzpicture}
	\begin{pgfonlayer}{nodelayer}
		\node [style=none] (16) at (3.5, 8) {};
		\node [style=none] (17) at (4, 7.5) {};
		\node [style=none] (18) at (4, 6.5) {};
		\node [style=none] (19) at (3.5, 6) {};
		\node [style=none] (20) at (-3.5, 8) {};
		\node [style=none] (21) at (-4, 7.5) {};
		\node [style=none] (22) at (-4, 6.5) {};
		\node [style=none] (23) at (-3.5, 6) {};
		\node [style=sLabel] (33) at (0, -10.875) {End};
		\node [style=sLabel] (34) at (0, 9.375) {Start};
		\node [style=none] (35) at (0, 9) {};
		\node [style=none] (36) at (0, 8) {};
		\node [style=none] (37) at (13, -3.75) {};
		\node [style=none] (38) at (13, -4.75) {};
		\node [style=none] (40) at (13, 2.75) {};
		\node [style=none] (41) at (2, -8) {};
		\node [style=none] (282) at (4, 7) {};
		\node [style=none] (283) at (5.75, 4.75) {};
		\node [style=none] (284) at (4, 3.75) {};
		\node [style=none] (285) at (5.75, 2.75) {};
		\node [style=none] (286) at (7.5, 3.75) {};
		\node [style=sLabel] (287) at (5.75, 4) {Flag};
		\node [style=sLabel] (288) at (5.75, 3.5) {raised};
		\node [style=sLabel] (289) at (8, 4.25) {Yes};
		\node [style=none] (290) at (1.5, 4.875) {};
		\node [style=none] (291) at (1.5, 2.625) {};
		\node [style=none] (292) at (-1.5, 2.625) {};
		\node [style=none] (293) at (-1.5, 4.875) {};
		\node [style=sLabel] (294) at (0, 3.125) {Correction};
		\node [style=none] (295) at (0, 1.5) {};
		\node [style=none] (296) at (9, 3.75) {};
		\node [style=sLabel] (297) at (3.5, 4.25) {No};
		\node [style=sLabel] (298) at (0, 3.75) {Decoder \&};
		\node [style=none] (299) at (4, 0.5) {};
		\node [style=none] (300) at (5.75, -1.75) {};
		\node [style=none] (301) at (4, -2.75) {};
		\node [style=none] (302) at (5.75, -3.75) {};
		\node [style=none] (303) at (7.5, -2.75) {};
		\node [style=sLabel] (304) at (5.75, -2.5) {Flag};
		\node [style=sLabel] (305) at (5.75, -3) {raised};
		\node [style=sLabel] (306) at (8, -2.25) {Yes};
		\node [style=none] (307) at (0, -6.75) {};
		\node [style=none] (308) at (9, -2.75) {};
		\node [style=sLabel] (309) at (3.5, -2.25) {No};
		\node [style=none] (310) at (0, -6.75) {};
		\node [style=none] (311) at (-2, -8) {};
		\node [style=none] (312) at (0, -9.25) {};
		\node [style=none] (313) at (2, -8) {};
		\node [style=sLabel] (314) at (0, -7.75) {Repeat};
		\node [style=sLabel] (315) at (0, -8.25) {QEC cycle};
		\node [style=sLabel] (316) at (-2.5, -7.5) {Yes};
		\node [style=sLabel] (317) at (0.5, -9.75) {No};
		\node [style=none] (318) at (0, -10.5) {};
		\node [style=none] (319) at (-5, -8) {};
		\node [style=none] (320) at (-5, 7) {};
		\node [style=none] (321) at (-4, 7) {};
		\node [style=sLabel] (322) at (0, 4.375) {Standard};
		\node [style=none] (323) at (1.5, -1.625) {};
		\node [style=none] (324) at (1.5, -3.875) {};
		\node [style=none] (325) at (-1.5, -3.875) {};
		\node [style=none] (326) at (-1.5, -1.625) {};
		\node [style=none] (327) at (14.5, 1.625) {};
		\node [style=none] (328) at (14.5, -0.625) {};
		\node [style=none] (329) at (11.5, -0.625) {};
		\node [style=none] (330) at (11.5, 1.625) {};
		\node [style=sLabel] (331) at (13, 1.125) {Modified};
		\node [style=none] (332) at (14.5, -4.75) {};
		\node [style=none] (333) at (14.5, -7) {};
		\node [style=none] (334) at (11.5, -7) {};
		\node [style=none] (335) at (11.5, -4.75) {};
		\node [style=sLabel] (336) at (13, -5.25) {Modified};
		\node [style=sLabel] (337) at (0, -3.375) {Correction};
		\node [style=sLabel] (338) at (0, -2.75) {Decoder \&};
		\node [style=sLabel] (339) at (0, -2.125) {Standard};
		\node [style=sLabel] (340) at (13, -0.125) {Correction};
		\node [style=sLabel] (341) at (13, 0.5) {Decoder \&};
		\node [style=sLabel] (342) at (13, -6.5) {Correction};
		\node [style=sLabel] (343) at (13, -5.875) {Decoder \&};
		\node [style=none] (344) at (5.75, 7) {};
		\node [style=none] (345) at (3.25, 3.75) {};
		\node [style=none] (346) at (1.5, 3.75) {};
		\node [style=none] (347) at (0, 2) {};
		\node [style=none] (348) at (5.75, 0.5) {};
		\node [style=none] (349) at (2, -2.75) {};
		\node [style=none] (350) at (2, -2.75) {};
		\node [style=none] (351) at (0, -6.75) {};
		\node [style=none] (352) at (1.5, -2.75) {};
		\node [style=none] (353) at (0, -3.875) {};
		\node [style=none] (355) at (14.5, -5.875) {};
		\node [style=none] (356) at (13, -8) {};
		\node [style=none] (357) at (13, -7) {};
		\node [style=none] (358) at (2, -8) {};
		\node [style=none] (359) at (13, 1.625) {};
		\node [style=none] (360) at (14.5, 0.5) {};
		\node [style=none] (361) at (17.5, 0.5) {};
		\node [style=none] (362) at (17.5, -8) {};
		\node [style=none] (364) at (0, 2.625) {};
		\node [style=none] (365) at (0, 7) {Primary Z-SE Circuit};
		\node [style=none] (366) at (3.5, 1.5) {};
		\node [style=none] (367) at (4, 1) {};
		\node [style=none] (368) at (4, 0) {};
		\node [style=none] (369) at (3.5, -0.5) {};
		\node [style=none] (370) at (-3.5, 1.5) {};
		\node [style=none] (371) at (-4, 1) {};
		\node [style=none] (372) at (-4, 0) {};
		\node [style=none] (373) at (-3.5, -0.5) {};
		\node [style=none] (374) at (0, 1.5) {};
		\node [style=none] (375) at (4, 0.5) {};
		\node [style=none] (376) at (-4, 0.5) {};
		\node [style=none] (377) at (0, 0.5) {Primary X-SE Circuit};
		\node [style=none] (378) at (16.5, 4.75) {};
		\node [style=none] (379) at (17, 4.25) {};
		\node [style=none] (380) at (17, 3.25) {};
		\node [style=none] (381) at (16.5, 2.75) {};
		\node [style=none] (382) at (9.5, 4.75) {};
		\node [style=none] (383) at (9, 4.25) {};
		\node [style=none] (384) at (9, 3.25) {};
		\node [style=none] (385) at (9.5, 2.75) {};
		\node [style=none] (386) at (13, 2.75) {};
		\node [style=none] (387) at (17, 3.75) {};
		\node [style=none] (388) at (9, 3.75) {};
		\node [style=none] (389) at (13, 3.75) {Recovery X-SE Circuit};
		\node [style=none] (390) at (16.5, -1.75) {};
		\node [style=none] (391) at (17, -2.25) {};
		\node [style=none] (392) at (17, -3.25) {};
		\node [style=none] (393) at (16.5, -3.75) {};
		\node [style=none] (394) at (9.5, -1.75) {};
		\node [style=none] (395) at (9, -2.25) {};
		\node [style=none] (396) at (9, -3.25) {};
		\node [style=none] (397) at (9.5, -3.75) {};
		\node [style=none] (398) at (13, -1.75) {};
		\node [style=none] (399) at (17, -2.75) {};
		\node [style=none] (400) at (9, -2.75) {};
		\node [style=none] (401) at (13, -2.75) {Recovery Z-SE Circuit};
	\end{pgfonlayer}
	\begin{pgfonlayer}{edgelayer}
		\draw [fill=verylightgray] (17.center)
			 to [bend right=45] (16.center)
			 to (20.center)
			 to [bend right=45] (21.center)
			 to (22.center)
			 to [bend right=45] (23.center)
			 to (19.center)
			 to [bend right=45] (18.center)
			 to cycle;
		\draw [style=arrow] (35.center) to (36.center);
		\draw [fill=yellow] (286.center)
			 to (285.center)
			 to (284.center)
			 to (283.center)
			 to cycle;
		\draw (292.center)
			 to (291.center)
			 to (290.center)
			 to (293.center)
			 to cycle;
		\draw [style=arrow] (286.center) to (296.center);
		\draw [fill=yellow] (303.center)
			 to (302.center)
			 to (301.center)
			 to (300.center)
			 to cycle;
		\draw [style=arrow] (303.center) to (308.center);
		\draw [fill=yellow] (313.center)
			 to (312.center)
			 to (311.center)
			 to (310.center)
			 to cycle;
		\draw (319.center) to (320.center);
		\draw [style=arrow] (320.center) to (321.center);
		\draw (325.center)
			 to (324.center)
			 to (323.center)
			 to (326.center)
			 to cycle;
		\draw (329.center)
			 to (328.center)
			 to (327.center)
			 to (330.center)
			 to cycle;
		\draw (334.center) to (333.center);
		\draw (333.center) to (332.center);
		\draw (332.center) to (335.center);
		\draw (335.center) to (334.center);
		\draw (344.center) to (282.center);
		\draw (345.center) to (284.center);
		\draw [style=arrow] (347.center) to (295.center);
		\draw [style=arrow] (344.center) to (283.center);
		\draw [style=arrow] (348.center) to (300.center);
		\draw (299.center) to (348.center);
		\draw (301.center) to (349.center);
		\draw (349.center) to (350.center);
		\draw [style=arrow] (350.center) to (352.center);
		\draw [style=arrow] (357.center) to (356.center);
		\draw [style=arrow] (312.center) to (318.center);
		\draw (311.center) to (319.center);
		\draw [style=arrow] (40.center) to (359.center);
		\draw (361.center) to (362.center);
		\draw [style=arrow] (362.center) to (41.center);
		\draw (360.center) to (361.center);
		\draw (364.center) to (347.center);
		\draw [style=arrow] (345.center) to (346.center);
		\draw [fill=verylightgray] (371.center)
			 to (372.center)
			 to [bend right=45] (373.center)
			 to (369.center)
			 to [bend right=45] (368.center)
			 to (367.center)
			 to [bend right=45] (366.center)
			 to (370.center)
			 to [bend right=45] cycle;
		\draw [fill=verylightgray] (383.center)
			 to (384.center)
			 to [bend right=45] (385.center)
			 to (381.center)
			 to [bend right=45] (380.center)
			 to (379.center)
			 to [bend right=45] (378.center)
			 to (382.center)
			 to [bend right=45] cycle;
		\draw [fill=verylightgray] (397.center)
			 to (393.center)
			 to [bend right=45] (392.center)
			 to (391.center)
			 to [bend right=45] (390.center)
			 to (394.center)
			 to [bend right=45] (395.center)
			 to (396.center)
			 to [bend right=45] cycle;
		\draw [style=arrow] (37.center) to (38.center);
		\draw [style=arrow] (353.center) to (351.center);
	\end{pgfonlayer}
\end{tikzpicture}}
  \]
  \caption{Flowchart of the full syndrome extraction protocol. A complete cycle consists of sequential Z and X syndrome extractions. For each basis, the protocol branches based on the primary circuit's flag outcome, routing to either standard decoding or the non-FT recovery circuit with modified decoding.}
  \label{fig:full-protocol}
\end{figure}

While the recovery circuit is not fault-tolerant on its own, we still obtain a protocol that is fault-tolerant according to the extended rectangle formalism for distance-three codes~\cite{aliferisQuantumAccuracy2006,gottesmanSurvivingQuantum2024}, following a similar argument to~\cite[Page 2]{chaoQuantumError2018}.

\paragraph{Fault-Tolerance via the Extended Rectangle Formalism:}
We define our fault model as the standard circuit-level depolarizing model (detailed further in \Cref{sec:simulations}).
Under this model, the protocol must handle at most a single fault.
Because the recovery circuit is only triggered when a flag has already isolated an internal fault in the primary circuit, it is guaranteed to operate on an input with bounded errors.
This allows it to fulfill the extended rectangle requirements without needing to be independently fault-tolerant.
Specifically:
\begin{enumerate}
  \item[(i)] if there is at most a single data error before measurement and no internal fault occurs, it appropriately corrects the error;
  \item[(ii)] if the data lies in the codespace and at most one internal fault occurs in the primary circuit, then
  \begin{enumerate}
    \item[(a)] if the flag is not raised, at most a weight-one error propagates to the data.
    \item[(b)] if a flag is raised, then the subsequent recovery circuit suffices to correct any possible error, assuming that no additional errors take place.
  \end{enumerate}
\end{enumerate}

\subsection{Decoding Logic}\label{subsec:decoding-logic}

To correctly infer the required corrections while accounting for potential internal faults, we use a specific decoding procedure.
Here, we summarize the logic in explicit steps depending on the state of the flag qubit.
The complete explicit decoding lookup table for both scenarios is provided in Appendix~\ref{sec:decoding-tables}.

\subsubsection{No Flag Raised}\label{subsubsec:no-flag-raised}

When the primary circuit's flag qubit measures $0$, no internal fault is detected, and the decoder proceeds as follows:

\begin{description}
  \item[Step 1] Obtain the raw measurement outcome, a $3$-bit string $b$, from the three ancillae.

  \item[Step 2] Compute the standard 3-bit syndrome $s$ using the circuit's effective parity-check matrix via $s = H_x' \cdot b$.\footnote{This matrix is derived from the parity-check matrix of \eqref{eq:parity-check} by keeping only columns 2, 3, and 5, as detailed in \protect\Cref{thm:FT-floq-steane}.}:
  \begin{equation}
    H_x' = H_z' =
    \begin{pmatrix}
      1 & 1 & 0 \\
      1 & 1 & 1 \\
      0 & 1 & 0 \\
    \end{pmatrix}
    \label{eq:effective-parity-check}
  \end{equation}

  \item[Step 3] Match $s$ to the corresponding column of the full parity-check matrix $H_x$ to identify the location of the single-qubit error:
  \begin{equation}
    H_x = H_z =
    \begin{pmatrix}
      1 & 1 & 1 & 1 & 0 & 0 & 0 \\
      0 & 1 & 1 & 0 & 1 & 1 & 0 \\
      0 & 0 & 1 & 1 & 0 & 1 & 1 \\
    \end{pmatrix}
    \label{eq:parity-check}
  \end{equation}
\end{description}
For example, a raw measurement of $b = (0, 1, 1)^T$ yields $s = (1, 0, 1)^T$, matching the 4th column of $H_x$ and indicating a correction on data qubit 4.

\subsubsection{Flag Raised}\label{subsubsec:flag-raised}

A flag outcome of $1$ indicates an internal fault in the primary circuit.
Because these faults can propagate into multi-qubit errors (e.g.\@ a specific weight-two error shown in \Cref{fig:example-error-propagation}), the standard decoder would misidentify them and introduce logical errors.
To prevent this, a modified decoding logic is applied to the results of the subsequent recovery circuit:

\begin{description}
  \item[Step 1] Obtain the new raw measurement outcome from the recovery circuit and calculate the new syndrome $s$.

  \item[Step 2] Apply corrections based on a dynamically remapped syndrome table.
  By examining the exhaustive list of flagged faults and their resulting data-qubit errors in \Cref{fig:detected-errors}, one can verify that only two specific faults result in dangerous weight-two errors ($Z_1 Z_2$ and $Z_2 Z_5$) that are not stabilizer-equivalent to a single-qubit error.
  To account for this, remap the syndromes as follows:
  \begin{itemize}
    \item If $s = (0, 1, 0)^T$, apply a correction to data qubits 1 and 2.
    \item If $s = (1, 0, 0)^T$, apply a correction to data qubits 2 and 5.
    \item For all other values of $s$, map to the standard single-qubit correction.
  \end{itemize}
\end{description}

Assuming only a single fault occurs, this remapping unambiguously corrects the specific weight-two data errors caused by internal faults, while properly handling standard weight-one errors.

\begin{figure}
  \begin{subfigure}{\linewidth}
    \[
      \begin{tikzpicture}
	\begin{pgfonlayer}{nodelayer}
		\node [style=not] (70) at (5, -2.75) {};
		\node [style=not] (71) at (5.75, -3.75) {};
		\node [style=not] (72) at (-5.5, -3.75) {};
		\node [style=not] (73) at (6.5, -4.75) {};
		\node [style=not] (74) at (-4.75, -4.75) {};
		\node [style=not] (75) at (-3.75, -2.75) {};
		\node [style=none] (76) at (-6.25, -3.75) {};
		\node [style=none] (77) at (-6.25, -4.75) {};
		\node [style=none] (78) at (-6.25, -2.75) {};
		\node [style=black dot] (79) at (1.5, 2) {};
		\node [style=black dot] (80) at (5, 1.375) {};
		\node [style=black dot] (81) at (5.75, 0.75) {};
		\node [style=black dot] (82) at (-5.5, 0.125) {};
		\node [style=black dot] (83) at (6.5, -0.5) {};
		\node [style=black dot] (84) at (-4.75, -1.125) {};
		\node [style=black dot] (85) at (-3.75, -1.75) {};
		\node [style=none] (86) at (7.75, 2) {};
		\node [style=none] (87) at (7.75, 1.375) {};
		\node [style=none] (88) at (7.75, 0.75) {};
		\node [style=none] (89) at (7.75, 0.125) {};
		\node [style=none] (90) at (7.75, -0.5) {};
		\node [style=none] (91) at (7.75, -1.125) {};
		\node [style=none] (92) at (7.75, -1.75) {};
		\node [style=none] (93) at (-6.25, 2) {};
		\node [style=none] (94) at (-6.25, 1.375) {};
		\node [style=none] (95) at (-6.25, 0.75) {};
		\node [style=none] (96) at (-6.25, 0.125) {};
		\node [style=none] (97) at (-6.25, -0.5) {};
		\node [style=none] (98) at (-6.25, -1.125) {};
		\node [style=none] (99) at (-6.25, -1.75) {};
		\node [style=black dot] (100) at (3, -2.75) {};
		\node [style=not] (101) at (3, -4.75) {};
		\node [style=black dot] (102) at (-1.5, -4.75) {};
		\node [style=not] (103) at (-1.5, -2.75) {};
		\node [style=black dot] (104) at (0, -2.75) {};
		\node [style=not] (105) at (1.5, -2.75) {};
		\node [style=not] (106) at (1.5, -4.75) {};
		\node [style=black dot] (107) at (1.5, -3.75) {};
		\node [style=not] (108) at (0, -3.75) {};
		\node [style=sLabel] (109) at (-6.75, -2.75) {$\ket{0}$};
		\node [style=sLabel] (110) at (-6.75, -3.75) {$\ket{0}$};
		\node [style=sLabel] (111) at (-6.75, -4.75) {$\ket{0}$};
		\node [style=meter] (115) at (7.5, -3.75) {};
		\node [style=meter] (116) at (7.5, -2.75) {};
		\node [style=meter] (117) at (7.5, -4.75) {};
		\node [style=not] (121) at (4, -2.75) {};
		\node [style=not] (122) at (-3.75, -4.75) {};
		\node [style=not] (123) at (-2.75, -3.75) {};
		\node [style=black dot] (124) at (-3.75, -5.75) {};
		\node [style=black dot] (125) at (-2.75, -5.75) {};
		\node [style=black dot] (126) at (4, -5.75) {};
		\node [style=meter] (127) at (7.5, -5.75) {};
		\node [style=none] (128) at (-6.25, -5.75) {};
		\node [style=sLabel] (129) at (-6.75, -5.75) {$\ket{0}$};
		\node [style=floatingLabel] (130) at (3.5, -2.375) {\tiny $\{2\}$};
		\node [style=floatingLabel] (131) at (2.25, -4.375) {\tiny $\{2, 5\}$};
		\node [style=floatingLabel] (132) at (2.25, -2.375) {\tiny $\{2\}$};
		\node [style=floatingLabel] (133) at (0.75, -2.375) {\tiny $\{1, 2\}$};
		\node [style=floatingLabel] (134) at (-0.75, -2.375) {\tiny $\{1, 2\}$};
		\node [style=floatingLabel] (135) at (-1.325, -3.375) {\tiny $\{1, 2, 3\}$};
		\node [style=floatingLabel] (136) at (0, -4.375) {\tiny $\{2, 3, 5\}$};
		\node [style=floatingLabel] (137) at (-2.575, -4.375) {\tiny $\{2, 3, 5\}$};
		\node [style=floatingLabel] (138) at (-4.25, -5.375) {\tiny $\{\}$};
		\node [style=floatingLabel] (139) at (-3.25, -5.375) {\tiny $\{\}$};
		\node [style=floatingLabel] (140) at (0, -5.375) {\tiny $\{\}$};
		\node [style=floatingLabel] (141) at (4.75, -5.375) {\tiny $\{\}$};
		\node [style=sLabel] (142) at (-6.75, 2) {1};
		\node [style=sLabel] (143) at (-6.75, 1.375) {2};
		\node [style=sLabel] (144) at (-6.75, 0.75) {3};
		\node [style=sLabel] (145) at (-6.75, 0.125) {4};
		\node [style=sLabel] (146) at (-6.75, -0.5) {5};
		\node [style=sLabel] (147) at (-6.75, -1.125) {6};
		\node [style=sLabel] (148) at (-6.75, -1.75) {7};
		\node [style=box] (149) at (5.75, -5.75) {$H$};
		\node [style=box] (150) at (-5, -5.75) {$H$};
	\end{pgfonlayer}
	\begin{pgfonlayer}{edgelayer}
		\draw (78.center) to (75);
		\draw (76.center) to (72);
		\draw (77.center) to (74);
		\draw (85) to (75);
		\draw (74) to (84);
		\draw (73) to (83);
		\draw (82) to (72);
		\draw (71) to (81);
		\draw (80) to (70);
		\draw (99.center) to (85);
		\draw (85) to (92.center);
		\draw (91.center) to (84);
		\draw (84) to (98.center);
		\draw (97.center) to (83);
		\draw (83) to (90.center);
		\draw (89.center) to (82);
		\draw (82) to (96.center);
		\draw (95.center) to (81);
		\draw (81) to (88.center);
		\draw (87.center) to (80);
		\draw (80) to (94.center);
		\draw (93.center) to (79);
		\draw (86.center) to (79);
		\draw (102) to (103);
		\draw (104) to (105);
		\draw (100) to (101);
		\draw (107) to (108);
		\draw (108) to (104);
		\draw (101) to (106);
		\draw (75) to (103);
		\draw (73) to (101);
		\draw (71) to (107);
		\draw (105) to (79);
		\draw (105) to (100);
		\draw (103) to (104);
		\draw (106) to (107);
		\draw (102) to (106);
		\draw (128.center) to (124);
		\draw (124) to (125);
		\draw (125) to (126);
		\draw (126) to (127);
		\draw (126) to (121);
		\draw (70) to (121);
		\draw (121) to (100);
		\draw (108) to (123);
		\draw (123) to (72);
		\draw (74) to (122);
		\draw (122) to (102);
		\draw (122) to (124);
		\draw (123) to (125);
		\draw [style=X Web] (126) to (125);
		\draw [style=X Web] (124) to (125);
		\draw [style=X Web] (122) to (124);
		\draw [style=X Web] (123) to (125);
		\draw [style=X Web] (122) to (102);
		\draw [style=X Web] (123) to (108);
		\draw [style=X Web] (108) to (104);
		\draw [style=X Web] (104) to (105);
		\draw [style=X Web] (105) to (100);
		\draw [style=X Web] (100) to (121);
		\draw [style=X Web] (121) to (126);
		\draw [style=X Web] (100) to (101);
		\draw [style=X Web] (101) to (106);
		\draw [style=X Web] (106) to (102);
		\draw [style=X Web] (102) to (103);
		\draw [style=X Web] (103) to (104);
		\draw (116) to (70);
		\draw (71) to (115);
		\draw (117) to (73);
		\draw [style=X Web] (149) to (126);
		\draw [style=X Web] (124) to (150);
		\draw [style=Z Web] (127) to (149);
		\draw [style=Z Web] (150) to (128.center);
	\end{pgfonlayer}
\end{tikzpicture}
    \]
    \caption{The detecting region of the FT circuit.
    A fault on a highlighted edge is detected by the flag qubit.
    Each fault-location label corresponds to its resulting data-qubit error.}
    \label{fig:detected-errors}
  \end{subfigure}
  \begin{subfigure}{\linewidth}
    \[
      \begin{tikzpicture}
	\begin{pgfonlayer}{nodelayer}
		\node [style=not] (0) at (4, -1) {};
		\node [style=not] (1) at (4.75, -2) {};
		\node [style=not] (2) at (-5.75, -2) {};
		\node [style=not] (3) at (5.5, -3) {};
		\node [style=not] (4) at (-5, -3) {};
		\node [style=not] (5) at (-4, -1) {};
		\node [style=none] (6) at (-6.5, -2) {};
		\node [style=none] (7) at (-6.5, -3) {};
		\node [style=none] (8) at (-6.5, -1) {};
		\node [style=black dot] (9) at (0, 3.75) {};
		\node [style=black dot] (10) at (4, 3.125) {};
		\node [style=black dot] (11) at (4.75, 2.5) {};
		\node [style=black dot] (12) at (-5.75, 1.875) {};
		\node [style=black dot] (13) at (5.5, 1.25) {};
		\node [style=black dot] (14) at (-5, 0.625) {};
		\node [style=black dot] (15) at (-4, 0) {};
		\node [style=none] (16) at (7, 3.75) {};
		\node [style=none] (17) at (7, 3.125) {};
		\node [style=none] (18) at (7, 2.5) {};
		\node [style=none] (19) at (7, 1.875) {};
		\node [style=none] (20) at (7, 1.25) {};
		\node [style=none] (21) at (7, 0.625) {};
		\node [style=none] (22) at (7, 0) {};
		\node [style=none] (23) at (-6.5, 3.75) {};
		\node [style=none] (24) at (-6.5, 3.125) {};
		\node [style=none] (25) at (-6.5, 2.5) {};
		\node [style=none] (26) at (-6.5, 1.875) {};
		\node [style=none] (27) at (-6.5, 1.25) {};
		\node [style=none] (28) at (-6.5, 0.625) {};
		\node [style=none] (29) at (-6.5, 0) {};
		\node [style=black dot] (30) at (2, -1) {};
		\node [style=not] (31) at (2, -3) {};
		\node [style=black dot] (32) at (-2, -3) {};
		\node [style=not] (33) at (-2, -1) {};
		\node [style=black dot] (34) at (-1, -1) {};
		\node [style=not] (35) at (0, -1) {};
		\node [style=not] (36) at (0, -3) {};
		\node [style=black dot] (37) at (0, -2) {};
		\node [style=not] (38) at (-1, -2) {};
		\node [style=sLabel] (39) at (-7, -1) {$\ket{0}$};
		\node [style=sLabel] (40) at (-7, -2) {$\ket{0}$};
		\node [style=sLabel] (41) at (-7, -3) {$\ket{0}$};
		\node [style=meter] (45) at (6.75, -2) {};
		\node [style=meter] (46) at (6.75, -1) {};
		\node [style=meter] (47) at (6.75, -4) {};
		\node [style=box] (49) at (5, -4) {$H$};
		\node [style=not] (51) at (3, -1) {};
		\node [style=not] (52) at (-4, -3) {};
		\node [style=not] (53) at (-3, -2) {};
		\node [style=black dot] (54) at (-4, -4) {};
		\node [style=black dot] (55) at (-3, -4) {};
		\node [style=black dot] (56) at (3, -4) {};
		\node [style=meter] (57) at (6.75, -3) {};
		\node [style=none] (58) at (-6.5, -4) {};
		\node [style=sLabel] (59) at (-7, -4) {$\ket{0}$};
		\node [style=fault-location-faulty, draw=green] (61) at (1, -3) {$Z$};
		\node [style=sLabel] (62) at (-7, 3.75) {1};
		\node [style=sLabel] (63) at (-7, 3.125) {2};
		\node [style=sLabel] (64) at (-7, 2.5) {3};
		\node [style=sLabel] (65) at (-7, 1.875) {4};
		\node [style=sLabel] (66) at (-7, 1.25) {5};
		\node [style=sLabel] (67) at (-7, 0.625) {6};
		\node [style=sLabel] (68) at (-7, 0) {7};
		\node [style=box] (69) at (-5.25, -4) {$H$};
	\end{pgfonlayer}
	\begin{pgfonlayer}{edgelayer}
		\draw (8.center) to (5);
		\draw (6.center) to (2);
		\draw (7.center) to (4);
		\draw (15) to (5);
		\draw (4) to (14);
		\draw (3) to (13);
		\draw (12) to (2);
		\draw (1) to (11);
		\draw (10) to (0);
		\draw (29.center) to (15);
		\draw (15) to (22.center);
		\draw (21.center) to (14);
		\draw (14) to (28.center);
		\draw (27.center) to (13);
		\draw (13) to (20.center);
		\draw (19.center) to (12);
		\draw (12) to (26.center);
		\draw (25.center) to (11);
		\draw (11) to (18.center);
		\draw (17.center) to (10);
		\draw (10) to (24.center);
		\draw (23.center) to (9);
		\draw (16.center) to (9);
		\draw (32) to (33);
		\draw (34) to (35);
		\draw (30) to (31);
		\draw (37) to (38);
		\draw (38) to (34);
		\draw (31) to (36);
		\draw (5) to (33);
		\draw (3) to (31);
		\draw (1) to (37);
		\draw (35) to (9);
		\draw (35) to (30);
		\draw (33) to (34);
		\draw (36) to (37);
		\draw (32) to (36);
		\draw (47) to (49);
		\draw (58.center) to (54);
		\draw (54) to (55);
		\draw (55) to (56);
		\draw (56) to (51);
		\draw (0) to (51);
		\draw (51) to (30);
		\draw (38) to (53);
		\draw (53) to (2);
		\draw (4) to (52);
		\draw (52) to (32);
		\draw (52) to (54);
		\draw (53) to (55);
		\draw [style=ZerrorPropagation] (31) to (30);
		\draw [style=ZerrorPropagation] (30) to (51);
		\draw [style=ZerrorPropagation] (31) to (61);
		\draw [style=ZerrorPropagation] (51) to (56);
		\draw [style=ZerrorPropagation] (51) to (0);
		\draw [style=ZerrorPropagation] (0) to (10);
		\draw [style=ZerrorPropagation] (31) to (3);
		\draw [style=ZerrorPropagation] (3) to (13);
		\draw [style=ZerrorPropagationArrow] (10) to (17.center);
		\draw [style=ZerrorPropagationArrow] (13) to (20.center);
		\draw [style=XerrorPropagationArrow] (49) to (47);
		\draw (0) to (46);
		\draw (1) to (45);
		\draw (3) to (57);
		\draw [style=ZerrorPropagationArrow] (3) to (57);
		\draw [style=ZerrorPropagationArrow] (0) to (46);
		\draw [style=ZerrorPropagationArrow] (56) to (49);
	\end{pgfonlayer}
\end{tikzpicture}
    \]
    \caption{An example of fault propagation.
    The green octagon with a $Z$ label indicates a single-qubit Z-error.
    The propagation of this fault is indicated with arrows, green for Z-error, and red for X-error.
    The fault propagates to qubits 2 and 5, and is detected by the flag measurement (bottom).}
    \label{fig:example-error-propagation}
  \end{subfigure}
  \caption{Error detection and fault propagation within the primary fault-tolerant syndrome extraction circuit.}
  \label{fig:FT-steane-decoder}
\end{figure}

\section{Simulations}\label{sec:simulations}

To validate the performance of our protocol, we conducted Monte Carlo simulations using Stim~\cite{gidneyStimFast2021}.
We compare our syndrome extraction protocol with the following methods:
\begin{itemize}
  \item The \emph{Three Qubit} method~\cite{reichardtFaulttolerantQuantum2020}, as the most established and experimentally demonstrated method.
  \item \emph{Steane}-style syndrome extraction~\cite{steaneActiveStabilization1997} with Goto's state preparation~\cite{gotoMinimizingResource2016}, as the basis of this work.
  \item \emph{Optimized Steane}-style syndrome extraction~\cite{rodatzFaultTolerance2025}, the direct predecessor of this work.
\end{itemize}

\paragraph{Noise model}
We assume circuit-level depolarizing noise with error probabilities $\pTwo$ for two-qubit gates, $\pSPAM$ for (only) measurement, and $\pMem$ for idle qubits.
A $Z$-type memory error is applied to all idling qubits, and non-interacting gates are assumed to execute in parallel.
We set $\pTwo = \pSPAM = \pPhis$ and $\pMem = 0.1\,\pPhis$.

This parameter regime and scaling ratio are characteristic of state-of-the-art atomic architectures, such as trapped-ion and neutral atom systems~\cite{bluvsteinLogicalQuantum2024, ryan-andersonRealizationRealTime2021}, where long qubit coherence times naturally result in idle error rates that are substantially lower than active gate infidelities.
We employ the standard circuit-level depolarizing channel to isolate the algorithmic efficiency of the protocols from hardware-specific noise peculiarities, providing a platform-agnostic baseline comparison~\cite{chaoQuantumError2018,reichardtFaulttolerantQuantum2020}.
The relative performance improvements of our protocol are expected to be robust across different noise regimes, provided the two-qubit gate and measurement infidelities remain the dominant source of error.

\begin{figure}[t!]
  \includegraphics[width=\linewidth]{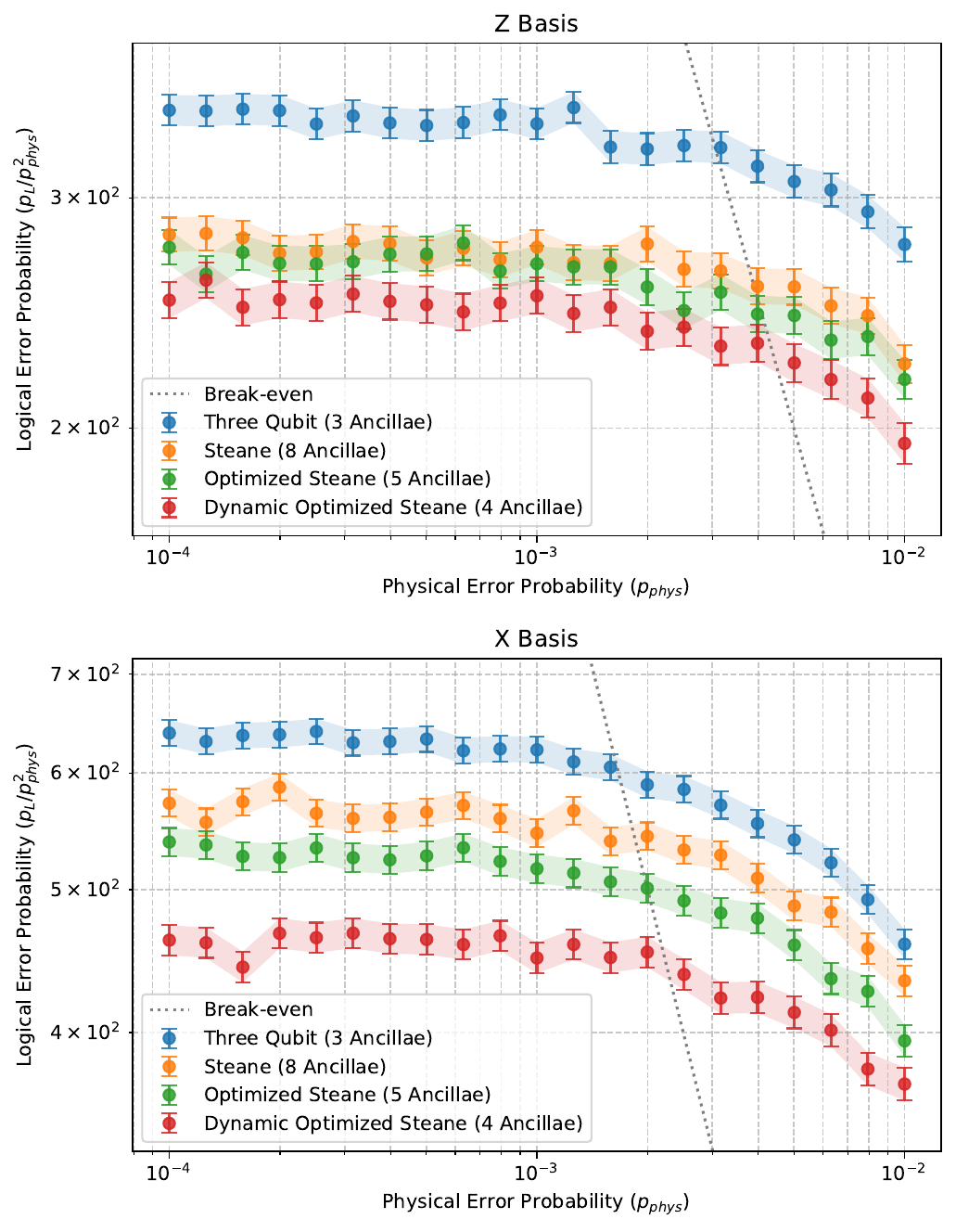}
  \caption{
    Logical error rate $\pL / p_{\mathrm{phys}}^2$ as a function of physical error rate $p_{\mathrm{phys}}$, with Wilson confidence intervals.}
  \label{subfig:le_per_pe}
\end{figure}

\paragraph{Results}
In the first set of simulations, we vary the physical error rate $\pPhis$ and compute the corresponding logical error probability $\pL$ per correction cycle.
Our method achieves a consistently lower logical error probability than previous approaches;
see \Cref{subfig:le_per_pe}.
To ensure a consistent number of logical errors and a stable confidence interval, we take $15 / \pPhis^2$ samples for each point.

\begin{figure}[t!]
  \includegraphics[width=\linewidth]{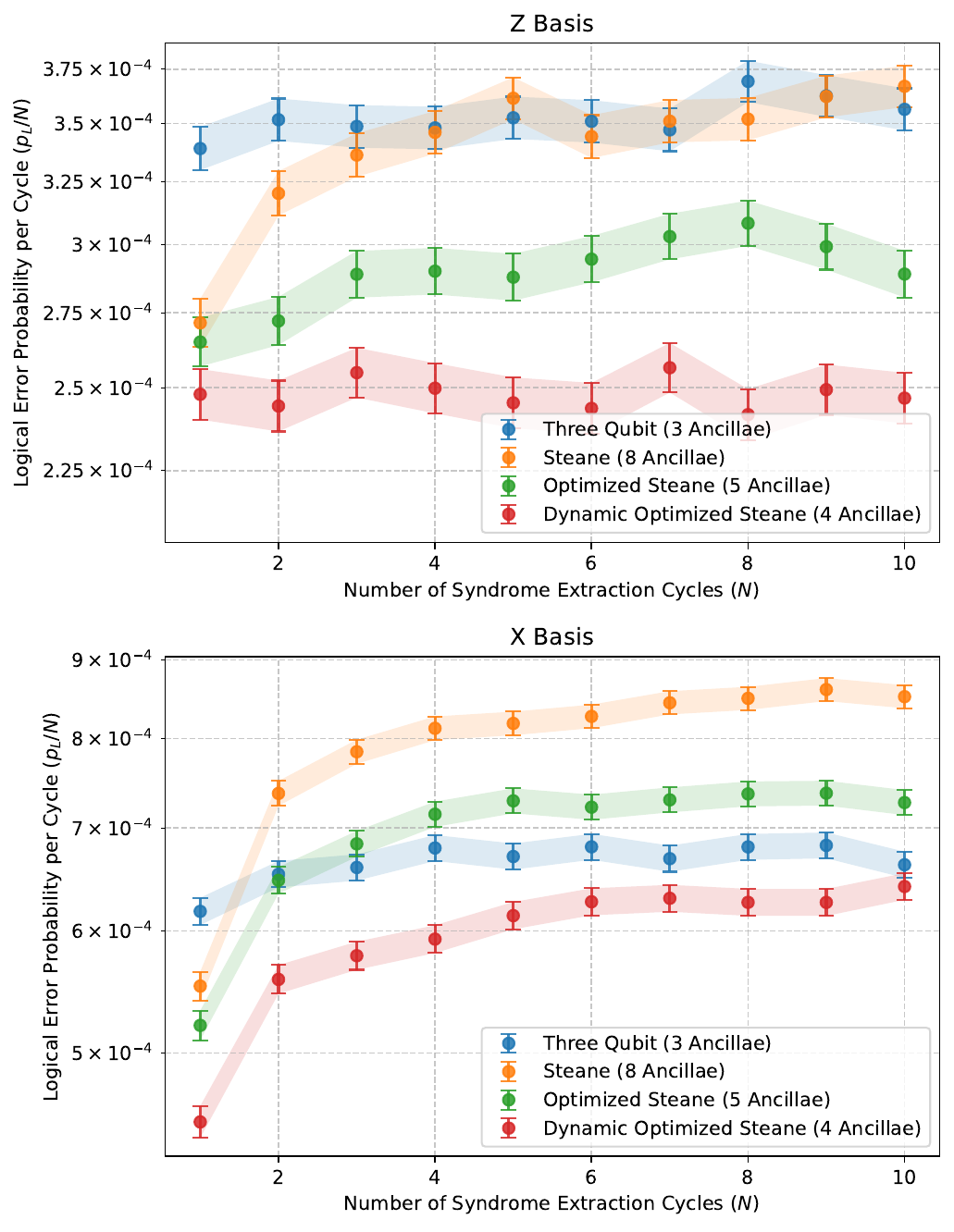}
  \caption{
    Logical error rate per cycles $\pL / N$ as a function of the number of error correction cycles $N$, with Wilson confidence intervals.}
  \label{subfig:le_per_cycles}
\end{figure}

In the second simulation, we evaluate performance under a realistic near-term hardware noise model with fixed parameters $\pTwo = \pSPAM = 10^{-3}$ and $\pMem = 10^{-4}$.
We repeat the syndrome measurement and correction cycles ($N$) and observe that our method consistently maintains the lowest logical error rates (\Cref{subfig:le_per_cycles}).
Each data point is averaged over $\frac{\pPhis^2}{15N}$ samples to obtain a consistent number of logical errors and a stable confidence interval.

\Cref{tab:avg-decrease} summarizes the mean relative reduction in logical error probability achieved by our protocol compared to prior methods.
These averages are calculated over the full sweep of simulated physical error rates (for the first experiment) and over all tested cycle counts (for the second experiment), aggregating the results from both the X and Z bases.

\begin{table}[t!]
  \centering
  \scriptsize
  \begin{tblr}{
    width = \linewidth,
    colspec = {X[0.7,c] l X[c] X[1.2,c] X[c]}, 
    rowsep = 2pt,
    stretch = 1,
    cells = {font=\small}
  }
    \toprule
    & & {vs.\@ Steane} & {vs.\@ Opt.\@ Steane} & {vs.\@ 3-qubit} \\
    \midrule
    \SetCell[r=3]{c}{\rotatebox{90}{\makecell[c]{\scriptsize Decrease\\ \scriptsize of $\frac{\pL}{\pPhis^2}$}}}
      & $Z$ basis  & 10.1\%  & 7.2\%  & 27.7\% \\
      & $X$ basis  & 17.8\% & 11.6\% & 25.5\% \\
      & Average    & 15.3\% & 10.1\% & 26.3\% \\
    \midrule
    \SetCell[r=3]{c}{\rotatebox{90}{\makecell[c]{\scriptsize Decrease\\ \scriptsize of $\frac{\pL}{N}$}}}
      & $Z$ basis & 26.8\% & 14.3\% & 29.6\% \\
      & $X$ basis & 24.9\% & 14.4\% & 10.8\% \\
      & Average   & 25.4\% & 14.3\% & 17.3\% \\
    \bottomrule
  \end{tblr}
  \caption{
    Average decrease (\%) in logical error probability of our method relative to prior approaches.
  }
  \label{tab:avg-decrease}
\end{table}

\paragraph{Expected overhead}

Finally, we compare the ancilla footprint, expected CNOT count, circuit depth, and overall logical performance of our protocol against prior methods.
The results are summarized in \Cref{tab:avg-stats}.

Because our method is adaptive, the required resources vary depending on whether a flag is raised.
The metrics reported for the Dynamic Optimized Steane method are therefore expected values.
A standard fault-free cycle requires 28 CNOTs (14 for the primary Z extraction and 14 for the primary X extraction).
If an internal fault triggers a flag, the protocol diverts to the 11-CNOT recovery circuit.

If the flag occurs during the first extraction, the cycle terminates early with a total of $14 + 11 = 25$ CNOTs.
If it occurs during the second extraction, the cycle requires $14 + 14 + 11 = 39$ CNOTs.
Thus, the expected number of CNOTs per cycle is given by:
\begin{equation}
  E[\text{CNOTs}] = 28 - 3p_1 + 11p_2
\end{equation}
where $p_1$ and $p_2$ are the empirical probabilities of a flag being raised in the first and second extractions, respectively.
Similarly, the expected depth is:
\begin{equation}
  E[\text{Depth}] = 20 - 2p_1 + 8p_2
\end{equation}

Following a similar reasoning, the expected number of CNOT gates and depth for the three-qubit method are calculated as follows:
\begin{gather}
  E[\text{CNOTs}] = 28 + 10p_1 + 24p_2 \\
  E[\text{Depth}] = 16 + 4p_1 + 12p_2
\end{gather}
where $p_1$ and $p_2$ are the empirical probabilities of a non-trivial measurement in the first and second extractions, respectively.

\begin{table*}[hbt!]
  \centering
  \scriptsize
  \begin{tblr}{
    width = \linewidth,
    colspec = {l X[c] X[c] X[1.2,c] X[1.4,c]}, 
    rowsep = 2pt,
    stretch = 1,
    cells = {font=\small}
  }
    \toprule
    & {Num.\@ Ancilla} & {Exp.\@ Num.\@ CNOTs} & {Exp.\@ Depth} & {Avg.\@ Error Prob.\@ / Cycle } \\
    \midrule
      Three qubit              & 3 & 29.42 & 16.67 & 0.283\% \\
      Steane                   & 8 & 36    & 20    & 0.325\% \\
      Optimized Steane         & 5 & 30    & 20    & 0.279\% \\
      Dynamic Optimized Steane & 4 & 28.08 & 20.06 & 0.238\% \\
    \bottomrule
  \end{tblr}
  \caption{
    Expected resources needed and average error rates for different methods based on the simulations with repeated error correction cycles.
    Statistics from left to right: number of ancilla qubits required, expected number of CNOT gates, expected depth, and average error probability per cycle.
  }
  \label{tab:avg-stats}
\end{table*}

\section{Summary and Outlook}

This paper introduces an adaptive, ultra low overhead fault-tolerant syndrome extraction protocol for the Steane code.
The protocol uses two highly optimized circuits derived using fault-equivalent ZX rewrites, a fault-tolerant and a non-fault-tolerant SE circuit with provably optimal CNOT counts in their respective settings.
Together, these circuits form a dynamic `flag-and-fallback' protocol that corrects both data errors and internal faults with minimal overhead.
Achieving further optimization for a code as extensively studied as the Steane code highlights that even small, well-understood codes offer room for improvement, and that combining fault-equivalent rewrites with code-level insights can help us discover new highly efficient methods.

The present protocol provides a practical template for distance-three CSS codes.
The process begins with optimizing a Steane-style syndrome extraction circuit using fault-equivalent rewrites.
As long as flagged internal faults result in unique syndromes, the same dynamic strategy applies directly to other distance-three codes.
Two main challenges remain in generalizing this approach.
For larger SE gadgets, the optimization process becomes more complex with increased circuit depth with current techniques.
This motivates the development of systematic, automated methods for fault-tolerant circuit optimization, potentially involving new classes of fault-equivalent rewrites.
The second challenge concerns the dynamic aspect itself:
for higher-distance codes, switching from FT to non-FT circuits would not satisfy Gottesman's criteria for fault tolerance.
Extending the approach to this regime will require a dynamic mechanism with intermediate steps to preserve full fault tolerance.

\section*{Data Availability}

The simulation files, numerical data, processing scripts necessary to reproduce the figures in this work, and files to perform the exhaustive searches described in Appendix~\ref{sec:cnot-optimal} are available at~\cite{Boldar99SteaneUltraLowOverhead}.
\\

\begin{acknowledgments}
  We thank Andrey Boris Khesin and Razin A.\@ Shaikh for discussions on optimality of CNOT counts.
  We appreciate the helpful reviews and numerous suggestions for improvement from David Amaro and Selwyn Simsek.
  BP and AK are supported by the Engineering and Physical Sciences Research Council grant number EP/Z002230/1, ``(De)constructing quantum software (DeQS)''.
  BR thanks Simon Harrison for his generous support for the Wolfson Harrison UK Research Council Quantum Foundation Scholarship.
  The authors used automated language tools for editorial assistance.
\end{acknowledgments}

\bibliography{ZX-QEC}

\appendix

\onecolumngrid
\setlength{\jot}{15pt}
\allowdisplaybreaks

\section{Derivations}\label{sec:derivations}

\begin{theorem}
  The following transformation is fault-equivalent:
  \[
    \begin{tikzpicture}
	\begin{pgfonlayer}{nodelayer}
		\node [style=not] (0) at (-1, -2) {};
		\node [style=black dot] (2) at (-1, -5) {};
		\node [style=box] (3) at (-6, -5) {$H$};
		\node [style=none] (4) at (-6.75, -2) {};
		\node [style=box] (7) at (-6, -7) {$H$};
		\node [style=box] (8) at (-6, -1) {$H$};
		\node [style=none] (11) at (-6.75, -6) {};
		\node [style=none] (12) at (-6.75, -3) {};
		\node [style=meter] (14) at (4, -8) {};
		\node [style=none] (15) at (-6.75, -8) {};
		\node [style=none] (16) at (-6.75, -4) {};
		\node [style=not] (18) at (-3, -6) {};
		\node [style=black dot] (19) at (-3, -7) {};
		\node [style=not] (20) at (-2, -4) {};
		\node [style=black dot] (21) at (-2, -1) {};
		\node [style=not] (22) at (-3, -3) {};
		\node [style=black dot] (23) at (-3, -5) {};
		\node [style=not] (24) at (-4, -4) {};
		\node [style=black dot] (25) at (-4, -7) {};
		\node [style=not] (26) at (-5, -2) {};
		\node [style=black dot] (27) at (-5, -1) {};
		\node [style=not] (28) at (0, -3) {};
		\node [style=black dot] (29) at (0, -4) {};
		\node [style=not] (30) at (-5, -6) {};
		\node [style=black dot] (31) at (-5, -5) {};
		\node [style=not] (32) at (1, -8) {};
		\node [style=black dot] (33) at (1, -2) {};
		\node [style=not] (34) at (3, -8) {};
		\node [style=black dot] (35) at (3, -6) {};
		\node [style=not] (36) at (2, -8) {};
		\node [style=black dot] (37) at (2, -4) {};
		\node [style=sLabel] (38) at (-7.25, -2) {$\ket{0}$};
		\node [style=sLabel] (39) at (-7.25, -1) {$\ket{0}$};
		\node [style=sLabel] (40) at (-7.25, -3) {$\ket{0}$};
		\node [style=sLabel] (41) at (-7.25, -4) {$\ket{0}$};
		\node [style=sLabel] (42) at (-7.25, -5) {$\ket{0}$};
		\node [style=sLabel] (43) at (-7.25, -6) {$\ket{0}$};
		\node [style=sLabel] (44) at (-7.25, -7) {$\ket{0}$};
		\node [style=sLabel] (45) at (-7.25, -8) {$\ket{0}$};
		\node [style=none] (46) at (-6.75, -1) {};
		\node [style=none] (47) at (-6.75, -5) {};
		\node [style=none] (48) at (-6.75, -7) {};
		\node [style=meter] (56) at (7, -5) {};
		\node [style=meter] (57) at (7, -2) {};
		\node [style=meter] (58) at (7, -7) {};
		\node [style=meter] (59) at (7, -1) {};
		\node [style=meter] (60) at (7, -6) {};
		\node [style=meter] (61) at (7, -3) {};
		\node [style=meter] (62) at (7, -4) {};
		\node [style=black dot] (63) at (4, -5) {};
		\node [style=black dot] (64) at (2.5, -2) {};
		\node [style=black dot] (65) at (5, -7) {};
		\node [style=black dot] (66) at (2, -1) {};
		\node [style=black dot] (67) at (4.5, -6) {};
		\node [style=black dot] (68) at (3, -3) {};
		\node [style=black dot] (69) at (3.5, -4) {};
		\node [style=none] (77) at (7.25, 1.5) {};
		\node [style=none] (78) at (7.25, 3.75) {};
		\node [style=none] (79) at (7.25, 0) {};
		\node [style=none] (80) at (7.25, 4.5) {};
		\node [style=none] (81) at (7.25, 0.75) {};
		\node [style=none] (82) at (7.25, 3) {};
		\node [style=none] (83) at (7.25, 2.25) {};
		\node [style=none] (91) at (-7.5, 3) {};
		\node [style=none] (92) at (-7.5, 0.75) {};
		\node [style=none] (93) at (-7.5, 4.5) {};
		\node [style=none] (94) at (-7.5, 0) {};
		\node [style=none] (95) at (-7.5, 3.75) {};
		\node [style=none] (96) at (-7.5, 1.5) {};
		\node [style=none] (97) at (-7.5, 2.25) {};
		\node [style=not] (98) at (3, 3) {};
		\node [style=not] (99) at (4.5, 0.75) {};
		\node [style=not] (100) at (2, 4.5) {};
		\node [style=not] (101) at (5, 0) {};
		\node [style=not] (102) at (2.5, 3.75) {};
		\node [style=not] (103) at (4, 1.5) {};
		\node [style=not] (104) at (3.5, 2.25) {};
		\node [style=box] (108) at (6, -1) {$H$};
		\node [style=box] (112) at (6, -7) {$H$};
		\node [style=box] (113) at (6, -6) {$H$};
		\node [style=box] (114) at (6, -5) {$H$};
		\node [style=box] (115) at (6, -4) {$H$};
		\node [style=box] (116) at (6, -3) {$H$};
		\node [style=box] (117) at (6, -2) {$H$};
	\end{pgfonlayer}
	\begin{pgfonlayer}{edgelayer}
		\draw (2) to (3);
		\draw (15.center) to (14);
		\draw (26) to (27);
		\draw (30) to (31);
		\draw (24) to (25);
		\draw (18) to (19);
		\draw (22) to (23);
		\draw (20) to (21);
		\draw (0) to (2);
		\draw (28) to (29);
		\draw (32) to (33);
		\draw (36) to (37);
		\draw (35) to (34);
		\draw (48.center) to (7);
		\draw (3) to (47.center);
		\draw (46.center) to (8);
		\draw (66) to (59);
		\draw (64) to (57);
		\draw (61) to (68);
		\draw (69) to (62);
		\draw (56) to (63);
		\draw (67) to (60);
		\draw (58) to (65);
		\draw (63) to (2);
		\draw (4.center) to (64);
		\draw (8) to (66);
		\draw (12.center) to (68);
		\draw (16.center) to (69);
		\draw (67) to (11.center);
		\draw (7) to (65);
		\draw (101) to (94.center);
		\draw (99) to (92.center);
		\draw (96.center) to (103);
		\draw (104) to (97.center);
		\draw (91.center) to (98);
		\draw (102) to (95.center);
		\draw (93.center) to (100);
		\draw (66) to (100);
		\draw (64) to (102);
		\draw (98) to (68);
		\draw (69) to (104);
		\draw (103) to (63);
		\draw (67) to (99);
		\draw (101) to (65);
		\draw (100) to (80.center);
		\draw (102) to (78.center);
		\draw (82.center) to (98);
		\draw (104) to (83.center);
		\draw (77.center) to (103);
		\draw (99) to (81.center);
		\draw (79.center) to (101);
	\end{pgfonlayer}
\end{tikzpicture}
    \qquad\FaultEq\qquad
    \begin{tikzpicture}
	\begin{pgfonlayer}{nodelayer}
		\node [style=black dot] (0) at (3.75, -1) {};
		\node [style=black dot] (1) at (4.5, -2) {};
		\node [style=black dot] (2) at (-5, -2) {};
		\node [style=black dot] (3) at (5.25, -3) {};
		\node [style=black dot] (4) at (-4.25, -3) {};
		\node [style=black dot] (5) at (-3.25, -1) {};
		\node [style=none] (6) at (-6.75, -2) {};
		\node [style=none] (7) at (-6.75, -3) {};
		\node [style=none] (8) at (-6.75, -1) {};
		\node [style=not] (9) at (0.75, 4.5) {};
		\node [style=not] (10) at (3.75, 3.75) {};
		\node [style=not] (11) at (4.5, 3) {};
		\node [style=not] (12) at (-5, 2.25) {};
		\node [style=not] (13) at (5.25, 1.5) {};
		\node [style=not] (14) at (-4.25, 0.75) {};
		\node [style=not] (15) at (-3.25, 0) {};
		\node [style=none] (16) at (7.5, 4.5) {};
		\node [style=none] (17) at (7.5, 3.75) {};
		\node [style=none] (18) at (7.5, 3) {};
		\node [style=none] (19) at (7.5, 2.25) {};
		\node [style=none] (20) at (7.5, 1.5) {};
		\node [style=none] (21) at (7.5, 0.75) {};
		\node [style=none] (22) at (7.5, 0) {};
		\node [style=none] (23) at (-6.75, 4.5) {};
		\node [style=none] (24) at (-6.75, 3.75) {};
		\node [style=none] (25) at (-6.75, 3) {};
		\node [style=none] (26) at (-6.75, 2.25) {};
		\node [style=none] (27) at (-6.75, 1.5) {};
		\node [style=none] (28) at (-6.75, 0.75) {};
		\node [style=none] (29) at (-6.75, 0) {};
		\node [style=not] (30) at (1.75, -1) {};
		\node [style=black dot] (31) at (1.75, -3) {};
		\node [style=not] (32) at (-1.25, -3) {};
		\node [style=black dot] (33) at (-1.25, -1) {};
		\node [style=not] (34) at (-0.25, -1) {};
		\node [style=black dot] (35) at (0.75, -1) {};
		\node [style=black dot] (36) at (0.75, -3) {};
		\node [style=not] (37) at (0.75, -2) {};
		\node [style=black dot] (38) at (-0.25, -2) {};
		\node [style=sLabel] (39) at (-7.25, -1) {$\ket{0}$};
		\node [style=sLabel] (40) at (-7.25, -2) {$\ket{0}$};
		\node [style=sLabel] (41) at (-7.25, -3) {$\ket{0}$};
		\node [style=box] (42) at (-6, -1) {$H$};
		\node [style=box] (43) at (-6, -2) {$H$};
		\node [style=box] (44) at (-6, -3) {$H$};
		\node [style=meter] (45) at (7.25, -2) {};
		\node [style=meter] (46) at (7.25, -1) {};
		\node [style=meter] (47) at (7.25, -3) {};
		\node [style=box] (48) at (6.25, -2) {$H$};
		\node [style=box] (49) at (6.25, -3) {$H$};
		\node [style=box] (50) at (6.25, -1) {$H$};
		\node [style=black dot] (51) at (2.75, -1) {};
		\node [style=black dot] (52) at (-3.25, -3) {};
		\node [style=black dot] (53) at (-2.25, -2) {};
		\node [style=not] (54) at (-3.25, -4) {};
		\node [style=not] (55) at (-2.25, -4) {};
		\node [style=not] (56) at (2.75, -4) {};
		\node [style=meter] (57) at (7.25, -4) {};
		\node [style=none] (58) at (-6.75, -4) {};
		\node [style=sLabel] (59) at (-7.25, -4) {$\ket{0}$};
		\node [style=none] (60) at (7.5, -4.5) {};
	\end{pgfonlayer}
	\begin{pgfonlayer}{edgelayer}
		\draw (8.center) to (5);
		\draw (6.center) to (2);
		\draw (7.center) to (4);
		\draw (15) to (5);
		\draw (4) to (14);
		\draw (3) to (13);
		\draw (12) to (2);
		\draw (1) to (11);
		\draw (10) to (0);
		\draw (29.center) to (15);
		\draw (15) to (22.center);
		\draw (21.center) to (14);
		\draw (14) to (28.center);
		\draw (27.center) to (13);
		\draw (13) to (20.center);
		\draw (19.center) to (12);
		\draw (12) to (26.center);
		\draw (25.center) to (11);
		\draw (11) to (18.center);
		\draw (17.center) to (10);
		\draw (10) to (24.center);
		\draw (23.center) to (9);
		\draw (16.center) to (9);
		\draw (32) to (33);
		\draw (34) to (35);
		\draw (30) to (31);
		\draw (37) to (38);
		\draw (38) to (34);
		\draw (31) to (36);
		\draw (5) to (33);
		\draw (3) to (31);
		\draw (1) to (37);
		\draw (35) to (9);
		\draw (35) to (30);
		\draw (33) to (34);
		\draw (36) to (37);
		\draw (32) to (36);
		\draw (47) to (49);
		\draw (49) to (3);
		\draw (1) to (48);
		\draw (48) to (45);
		\draw (46) to (50);
		\draw (50) to (0);
		\draw (58.center) to (54);
		\draw (54) to (55);
		\draw (55) to (56);
		\draw (56) to (57);
		\draw (56) to (51);
		\draw (0) to (51);
		\draw (51) to (30);
		\draw (38) to (53);
		\draw (53) to (2);
		\draw (4) to (52);
		\draw (52) to (32);
		\draw (52) to (54);
		\draw (53) to (55);
	\end{pgfonlayer}
\end{tikzpicture}
  \]
  \label{thm:FT-floq-steane}
\end{theorem}
\begin{proof}
  The fault-equivalence of such a mapping is established for CSS codes in~\cite[Section 7]{rodatzFaultTolerance2025}.
  \begin{align*}
    &\scalebox{.73}{\input{figures/FT-floq-steane-X-lemma-proof-1.tikz}}\\
    &\scalebox{.73}{\input{figures/FT-floq-steane-X-lemma-proof-2.tikz}}\\
    &\scalebox{.73}{\input{figures/FT-floq-steane-X-lemma-proof-3.tikz}}
  \end{align*}
  The proof uses several ZX rewrite rules, including Color Change, Fusion, Identity, and the OCM (Only Connected Measurements) rule.
  The fault-equivalence of these rewrites is proven in~\cite{rodatzFaultTolerance2025,rodatzFloquetifyingStabiliser2024}.

  A key step, marked by $(*)$ in the proof, involves setting the phase of certain measurement spiders to zero, which is equivalent to post-selecting on their measurement outcomes.
  However, because these spiders are not extracted as measurements in the final diagram, the process yields a physically realizable and fully deterministic quantum circuit with no post-selection and fewer ancillae.
  Despite this reduction, the full syndrome $s$ can still be recovered using an effective parity-check matrix, $H_x'$, which is formed by keeping only the columns of the original matrix $H_x$ that correspond to the remaining measurements (in this case, columns 2, 3, and 5).
  We can verify that no syndrome information is lost, since the effective and full matrices have the same rank:
  \[
    \rank(H_x)
    \quad = \quad
    \rank
    \begin{pmatrix}
      1 & 1 & 1 & 1 & 0 & 0 & 0 \\
      0 & 1 & 1 & 0 & 1 & 1 & 0 \\
      0 & 0 & 1 & 1 & 0 & 1 & 1 \\
    \end{pmatrix}
    \quad = \ 3
    \qquad\qquad
    \rank(H_x')
    \quad = \quad
    \rank
    \begin{pmatrix}
      1 & 1 & 0 \\
      1 & 1 & 1 \\
      0 & 1 & 0 \\
    \end{pmatrix}
    \quad = \ 3
  \]
  As such, either circuit provides three bits of information: the commutation relationship of the error with the three $X$ stabilizers.
  However, the usual Steane-style syndrome extraction circuit does so from seven measurements while the optimized circuit extracts the three bits of information from three measurements.
\end{proof}

\begin{theorem}
  The following transformation is fault-equivalent:
  \[
    \begin{tikzpicture}
	\begin{pgfonlayer}{nodelayer}
		\node [style=not] (0) at (0, -2) {};
		\node [style=black dot] (1) at (0, -5) {};
		\node [style=box] (2) at (-5, -5) {$H$};
		\node [style=none] (3) at (-5.75, -2) {};
		\node [style=box] (4) at (-5, -7) {$H$};
		\node [style=box] (5) at (-5, -1) {$H$};
		\node [style=none] (6) at (-5.75, -6) {};
		\node [style=none] (7) at (-5.75, -3) {};
		\node [style=none] (10) at (-5.75, -4) {};
		\node [style=not] (11) at (-2, -6) {};
		\node [style=black dot] (12) at (-2, -7) {};
		\node [style=not] (13) at (-1, -4) {};
		\node [style=black dot] (14) at (-1, -1) {};
		\node [style=not] (15) at (-2, -3) {};
		\node [style=black dot] (16) at (-2, -5) {};
		\node [style=not] (17) at (-3, -4) {};
		\node [style=black dot] (18) at (-3, -7) {};
		\node [style=not] (19) at (-4, -2) {};
		\node [style=black dot] (20) at (-4, -1) {};
		\node [style=not] (21) at (1, -3) {};
		\node [style=black dot] (22) at (1, -4) {};
		\node [style=not] (23) at (-4, -6) {};
		\node [style=black dot] (24) at (-4, -5) {};
		\node [style=sLabel] (31) at (-6.25, -2) {$\ket{0}$};
		\node [style=sLabel] (32) at (-6.25, -1) {$\ket{0}$};
		\node [style=sLabel] (33) at (-6.25, -3) {$\ket{0}$};
		\node [style=sLabel] (34) at (-6.25, -4) {$\ket{0}$};
		\node [style=sLabel] (35) at (-6.25, -5) {$\ket{0}$};
		\node [style=sLabel] (36) at (-6.25, -6) {$\ket{0}$};
		\node [style=sLabel] (37) at (-6.25, -7) {$\ket{0}$};
		\node [style=none] (39) at (-5.75, -1) {};
		\node [style=none] (40) at (-5.75, -5) {};
		\node [style=none] (41) at (-5.75, -7) {};
		\node [style=meter] (42) at (6, -5) {};
		\node [style=meter] (43) at (6, -2) {};
		\node [style=meter] (44) at (6, -7) {};
		\node [style=meter] (45) at (6, -1) {};
		\node [style=meter] (46) at (6, -6) {};
		\node [style=meter] (47) at (6, -3) {};
		\node [style=meter] (48) at (6, -4) {};
		\node [style=black dot] (49) at (3, -5) {};
		\node [style=black dot] (50) at (1.5, -2) {};
		\node [style=black dot] (51) at (4, -7) {};
		\node [style=black dot] (52) at (1, -1) {};
		\node [style=black dot] (53) at (3.5, -6) {};
		\node [style=black dot] (54) at (2, -3) {};
		\node [style=black dot] (55) at (2.5, -4) {};
		\node [style=none] (56) at (6.25, 1.5) {};
		\node [style=none] (57) at (6.25, 3.75) {};
		\node [style=none] (58) at (6.25, 0) {};
		\node [style=none] (59) at (6.25, 4.5) {};
		\node [style=none] (60) at (6.25, 0.75) {};
		\node [style=none] (61) at (6.25, 3) {};
		\node [style=none] (62) at (6.25, 2.25) {};
		\node [style=none] (63) at (-6.5, 3) {};
		\node [style=none] (64) at (-6.5, 0.75) {};
		\node [style=none] (65) at (-6.5, 4.5) {};
		\node [style=none] (66) at (-6.5, 0) {};
		\node [style=none] (67) at (-6.5, 3.75) {};
		\node [style=none] (68) at (-6.5, 1.5) {};
		\node [style=none] (69) at (-6.5, 2.25) {};
		\node [style=not] (70) at (2, 3) {};
		\node [style=not] (71) at (3.5, 0.75) {};
		\node [style=not] (72) at (1, 4.5) {};
		\node [style=not] (73) at (4, 0) {};
		\node [style=not] (74) at (1.5, 3.75) {};
		\node [style=not] (75) at (3, 1.5) {};
		\node [style=not] (76) at (2.5, 2.25) {};
		\node [style=box] (77) at (5, -1) {$H$};
		\node [style=box] (78) at (5, -7) {$H$};
		\node [style=box] (79) at (5, -6) {$H$};
		\node [style=box] (80) at (5, -5) {$H$};
		\node [style=box] (81) at (5, -4) {$H$};
		\node [style=box] (82) at (5, -3) {$H$};
		\node [style=box] (83) at (5, -2) {$H$};
	\end{pgfonlayer}
	\begin{pgfonlayer}{edgelayer}
		\draw (1) to (2);
		\draw (19) to (20);
		\draw (23) to (24);
		\draw (17) to (18);
		\draw (11) to (12);
		\draw (15) to (16);
		\draw (13) to (14);
		\draw (0) to (1);
		\draw (21) to (22);
		\draw (41.center) to (4);
		\draw (2) to (40.center);
		\draw (39.center) to (5);
		\draw (52) to (45);
		\draw (50) to (43);
		\draw (47) to (54);
		\draw (55) to (48);
		\draw (42) to (49);
		\draw (53) to (46);
		\draw (44) to (51);
		\draw (49) to (1);
		\draw (3.center) to (50);
		\draw (5) to (52);
		\draw (7.center) to (54);
		\draw (10.center) to (55);
		\draw (53) to (6.center);
		\draw (4) to (51);
		\draw (73) to (66.center);
		\draw (71) to (64.center);
		\draw (68.center) to (75);
		\draw (76) to (69.center);
		\draw (63.center) to (70);
		\draw (74) to (67.center);
		\draw (65.center) to (72);
		\draw (52) to (72);
		\draw (50) to (74);
		\draw (70) to (54);
		\draw (55) to (76);
		\draw (75) to (49);
		\draw (53) to (71);
		\draw (73) to (51);
		\draw (72) to (59.center);
		\draw (74) to (57.center);
		\draw (61.center) to (70);
		\draw (76) to (62.center);
		\draw (56.center) to (75);
		\draw (71) to (60.center);
		\draw (58.center) to (73);
	\end{pgfonlayer}
\end{tikzpicture}
    \qquad\FaultEq\qquad
    \begin{tikzpicture}
	\begin{pgfonlayer}{nodelayer}
		\node [style=black dot] (64) at (2.25, -1) {};
		\node [style=black dot] (65) at (3, -2) {};
		\node [style=black dot] (66) at (-4.25, -2) {};
		\node [style=black dot] (67) at (3.75, -3) {};
		\node [style=black dot] (68) at (-3.5, -3) {};
		\node [style=black dot] (69) at (-2.75, -1) {};
		\node [style=none] (72) at (-6, -2) {};
		\node [style=none] (74) at (-6, -3) {};
		\node [style=none] (75) at (-6, -1) {};
		\node [style=not] (76) at (0.25, 4.5) {};
		\node [style=not] (77) at (2.25, 3.75) {};
		\node [style=not] (78) at (3, 3) {};
		\node [style=not] (79) at (-4.25, 2.25) {};
		\node [style=not] (80) at (3.75, 1.5) {};
		\node [style=not] (81) at (-3.5, 0.75) {};
		\node [style=not] (82) at (-2.75, 0) {};
		\node [style=none] (83) at (6, 4.5) {};
		\node [style=none] (84) at (6, 3.75) {};
		\node [style=none] (85) at (6, 3) {};
		\node [style=none] (86) at (6, 2.25) {};
		\node [style=none] (87) at (6, 1.5) {};
		\node [style=none] (88) at (6, 0.75) {};
		\node [style=none] (89) at (6, 0) {};
		\node [style=none] (90) at (-6, 4.5) {};
		\node [style=none] (91) at (-6, 3.75) {};
		\node [style=none] (92) at (-6, 3) {};
		\node [style=none] (93) at (-6, 2.25) {};
		\node [style=none] (94) at (-6, 1.5) {};
		\node [style=none] (95) at (-6, 0.75) {};
		\node [style=none] (96) at (-6, 0) {};
		\node [style=not] (97) at (1.25, -1) {};
		\node [style=black dot] (98) at (1.25, -3) {};
		\node [style=not] (99) at (-1.75, -3) {};
		\node [style=black dot] (100) at (-1.75, -1) {};
		\node [style=not] (101) at (-0.75, -1) {};
		\node [style=black dot] (102) at (0.25, -1) {};
		\node [style=black dot] (103) at (0.25, -3) {};
		\node [style=not] (104) at (0.25, -2) {};
		\node [style=black dot] (105) at (-0.75, -2) {};
		\node [style=sLabel] (106) at (-6.5, -1) {$\ket{0}$};
		\node [style=sLabel] (107) at (-6.5, -2) {$\ket{0}$};
		\node [style=sLabel] (108) at (-6.5, -3) {$\ket{0}$};
		\node [style=box] (109) at (-5.25, -1) {$H$};
		\node [style=box] (112) at (-5.25, -2) {$H$};
		\node [style=box] (113) at (-5.25, -3) {$H$};
		\node [style=meter] (114) at (5.75, -2) {};
		\node [style=meter] (115) at (5.75, -1) {};
		\node [style=meter] (116) at (5.75, -3) {};
		\node [style=box] (117) at (4.75, -2) {$H$};
		\node [style=box] (118) at (4.75, -3) {$H$};
		\node [style=box] (119) at (4.75, -1) {$H$};
	\end{pgfonlayer}
	\begin{pgfonlayer}{edgelayer}
		\draw (75.center) to (69);
		\draw (72.center) to (66);
		\draw (74.center) to (68);
		\draw (82) to (69);
		\draw (68) to (81);
		\draw (67) to (80);
		\draw (79) to (66);
		\draw (65) to (78);
		\draw (77) to (64);
		\draw (96.center) to (82);
		\draw (82) to (89.center);
		\draw (88.center) to (81);
		\draw (81) to (95.center);
		\draw (94.center) to (80);
		\draw (80) to (87.center);
		\draw (86.center) to (79);
		\draw (79) to (93.center);
		\draw (92.center) to (78);
		\draw (78) to (85.center);
		\draw (84.center) to (77);
		\draw (77) to (91.center);
		\draw (90.center) to (76);
		\draw (83.center) to (76);
		\draw (99) to (100);
		\draw (101) to (102);
		\draw (97) to (98);
		\draw (104) to (105);
		\draw (105) to (101);
		\draw (98) to (103);
		\draw (69) to (100);
		\draw (67) to (98);
		\draw (65) to (104);
		\draw (97) to (64);
		\draw (105) to (66);
		\draw (68) to (99);
		\draw (102) to (76);
		\draw (102) to (97);
		\draw (100) to (101);
		\draw (103) to (104);
		\draw (99) to (103);
		\draw (116) to (118);
		\draw (118) to (67);
		\draw (65) to (117);
		\draw (117) to (114);
		\draw (115) to (119);
		\draw (119) to (64);
	\end{pgfonlayer}
\end{tikzpicture}
  \]
  \label{thm:non-FT-floq-steane}
\end{theorem}
\begin{proof}
  Same arguments as \Cref{thm:FT-floq-steane}.
  \begin{align*}
    &\scalebox{.8}{\input{figures/non-FT-floq-steane-X-lemma-proof-1.tikz}}\\
    &\scalebox{.8}{\input{figures/non-FT-floq-steane-X-lemma-proof-2.tikz}}\\
    &\scalebox{.8}{\input{figures/non-FT-floq-steane-X-lemma-proof-3.tikz}}
  \end{align*}
  Note: While we only use fault-equivalent rewrites and show that the transformation is fault-equivalent, it is not necessary for this proof as it is already a non-FT implementation of the SE circuit.
\end{proof}

\section{Optimality of CNOT count}
\label{sec:cnot-optimal}

For the task of non-destructive syndrome measurement of the Steane code, we argue that our methods are optimal.
First, we show that the non-FT circuit is optimal for CNOT count, and then argue that flagging such an implementation requires at least 3 CNOTs, making the FT version also optimal in CNOT count.

\NonFtCnotOptimality*
\begin{proof}\label{proof:non-ft-cnot}
  This problem can be understood as the construction of the parity-check matrix where only certain operations are allowed.
  Model the state of the ancilla qubits by a $3\times7$ binary matrix $M$ whose rows index ancillae $a_0,a_1,a_2$ and columns index data qubits $d_0,\dots,d_6$.
  The initial matrix is the zero matrix and the target is the Steane parity-check matrix
  \[
    H
    =
    \begin{pmatrix}
      1 & 1 & 1 & 1 & 0 & 0 & 0 \\
      0 & 1 & 1 & 0 & 1 & 1 & 0 \\
      0 & 0 & 1 & 1 & 0 & 1 & 1
    \end{pmatrix}.
  \]
  Each allowed primitive operation counts for one CNOT:
  \begin{itemize}
    \item a \emph{data$\to$ancilla} CNOT toggles a single entry $M_{ij}$ (a single-entry flip);
    \item an \emph{ancilla$\to$ancilla} CNOT adds one row to another (row$_s\gets$row$_s+$row$_r$).
  \end{itemize}

  \paragraph*{Proof by exhaustion (computer-assisted)}
  As the state space of $3\times7$ binary matrices is finite ($2^{21}$ states) and the allowed moves are simple, we can view the process as a finite unweighted directed graph whose vertices are $3\times7$ binary matrices and whose edges correspond to the primitive operations.
  A breadth-first search (BFS) from the zero matrix visits vertices in order of increasing total number of primitive operations;
  the distance (graph shortest-path length) from the zero matrix to $H$ is therefore the minimal number of CNOTs required in this setting.

  We performed an exhaustive BFS from the zero matrix and found that no path of length $\le 10$ reaches $H$, while there exists a path of length $11$, and indeed, \Cref{fig:non-FT-circuit} is one such construction.
  Therefore, the graph distance is exactly $11$, and therefore any protocol of the allowed type requires at least $11$ CNOTs.

  Implementation of the BFS can be found in~\cite{Boldar99SteaneUltraLowOverhead}.
\end{proof}

\FtCnotOptimality*
\begin{proof}\label{proof:ft-cnot}
  We give a computer-assisted exhaustive proof in the same style as the non-fault-tolerant optimality theorem.
  We consider all length-11 circuits that measure the three $Z$-syndromes using the allowed operations.
  For every such circuit $C$, in order to flag every single-ancilla $Z$ fault that could propagate to a weight-two $Z$ on the data, one must add at least three extra CNOT gates to the circuit.
  The proof proceeds by exhaustive enumeration and exact Pauli-propagation simulation.

  \paragraph*{Proof by exhaustion (computer-assisted)}
  The search space of base circuits was exhaustively enumerated.
  For each base circuit $C$, the following steps were performed:
  \begin{enumerate}
    \item Simulate Pauli propagation to list all ancilla space-time locations at which an ancilla-$Z$ fault would propagate into a weight-two data $Z$, up to multiplication by stabilizers;
    \item Search all sets of additional CNOTs that can flag any single fault identified in (1);
    \item Record the minimal number $m(C)$ of extra CNOTs required.
  \end{enumerate}

  \paragraph*{Result of the exhaustive search}
  The exhaustive run found that, for every enumerated and valid length-11 circuit, the value $m(C)$ satisfies $m(C)\ge 3$.
  In other words, with one or two additional CNOTs, no length-11 implementation can flag all ancilla-$Z$ faults that would otherwise propagate to weight-two data errors.
  Hence, to convert any such 11-CNOT implementation into one that flags all dangerous ancilla faults on a separate ancilla (i.e.\@ so that each dangerous fault raises a flag), one must add at least three additional CNOT gates.

  Implementation of the exhaustive search and a summary of the run can be found in~\cite{Boldar99SteaneUltraLowOverhead}.
\end{proof}

\section{Decoding Tables}\label{sec:decoding-tables}

\Cref{tab:decoding-lookup} provides the explicit syndrome-to-correction mapping used in our protocol.
The syndromes $s = (s_1, s_2, s_3)^T$ are computed from the raw measurement bits as described in \Cref{subsec:decoding-logic}.
The standard mapping corrects single-qubit errors when no flag is raised.
The modified mapping safely handles the specific weight-two data errors ($Z_1 Z_2$ and $Z_2 Z_5$) that can arise from internal faults in our primary extraction circuit when a flag is raised.

\begin{table}[h!]
  \centering
  \begin{tblr}{
    colspec = {c c c},
    rowsep = 3pt,
    cells = {font=\small},
    row{1} = {font=\bfseries}
  }
    \toprule
    Syndrome $s = (s_1, s_2, s_3)^T$ & Standard Correction (No Flag) & Modified Correction (Flag Raised) \\
    \midrule
    $(0, 0, 0)^T$ & None & None \\
    $(0, 0, 1)^T$ & Qubit 7 & Qubit 7 \\
    $(0, 1, 0)^T$ & Qubit 5 & Qubits 1 and 2 \\
    $(0, 1, 1)^T$ & Qubit 6 & Qubit 6 \\
    $(1, 0, 0)^T$ & Qubit 1 & Qubits 2 and 5 \\
    $(1, 0, 1)^T$ & Qubit 4 & Qubit 4 \\
    $(1, 1, 0)^T$ & Qubit 2 & Qubit 2 \\
    $(1, 1, 1)^T$ & Qubit 3 & Qubit 3 \\
    \bottomrule
  \end{tblr}
  \caption{Lookup table for the dynamic decoding logic. Data qubit indices correspond to the layout shown in Figure~1. Depending on whether an $X$ or $Z$ syndrome was measured, the appropriate Pauli $Z$ or $X$ corrections are applied to the listed qubits.}
  \label{tab:decoding-lookup}
\end{table}

\end{document}